\documentclass{emulateapj}
\usepackage{times}
\usepackage{lscape}

\def\ga{\mathrel{\mathpalette\fun >}}

\def\fun#1#2{\lower0.837ex\vbox{\baselineskip0ex\lineskip0.209ex
  \ialign{$\mathsurround=0ex#1\hfil##\hfil$\crcr#2\crcr\sim\crcr}}}

\def\degr{^\circ}

\def\sles{\lower2pt\hbox{$\buildrel {\scriptstyle <}
   \over {\scriptstyle\sim}$}}

\def\sgreat{\lower2pt\hbox{$\buildrel {\scriptstyle >}
   \over {\scriptstyle\sim}$}}

\def\ga{\mathrel{\mathpalette\fun >}}

\def\gks{}

\begin{document}

\title{{\it Swift} BAT Survey of 
AGN}
\author{
J.~Tueller\altaffilmark{1},
R.~F.~Mushotzky\altaffilmark{1},
S.~Barthelmy\altaffilmark{1},
J.~K.~Cannizzo\altaffilmark{1,2},
N.~Gehrels\altaffilmark{1},
C.~B.~Markwardt\altaffilmark{1,3},
G.~K. ~Skinner\altaffilmark{1,3},
L.~M.~Winter\altaffilmark{1,3}
}
\altaffiltext{1}{NASA/Goddard Space Flight Center, Astrophysics Science Division, Greenbelt, MD 20771}
\altaffiltext{2}{CRESST/Joint Center for Astrophysics, University of Maryland, Baltimore County,
                 Baltimore, MD 21250}
\altaffiltext{3}{CRESST/Department of Astronomy, University of Maryland College Park,
                 College Park, MD 20742}

\begin{abstract}
We present the results\footnote{to appear in The Astrophysical Journal, July 10, 2008, v. 681}
 of the analysis of the first 9 months of data of the
{\it Swift} BAT survey of AGN in the $14-195$ keV band. 
 Using archival X-ray data or follow-up {\it Swift} XRT observations, we  have identified 
129  (103 AGN) of 130
 objects  detected at $|b|> 15\degr$ and with significance $>4.8\sigma$. 
One source remains  unidentified.
These same X-ray data have allowed measurement
of the X-ray properties of the objects. We fit a power law to the $\log N - \log S$ distribution,
and find the slope to be $1.42\pm0.14$. 
Characterizing  the differential luminosity function data as   a broken power  law,  we find
a break luminosity $\log L_*$(erg s$^{-1}$)$=43.85\pm0.26         $, a low luminosity power law slope
$a=0.84^{+0.16}_{-0.22}$, 
                  and a high luminosity power law slope  $b=2.55^{+0.43}_{-0.30}$,
similar to the values  that have been reported  based on {\it INTEGRAL} data.
We obtain a mean photon index $1.98$ in the $14-195$ keV band, with an {\it rms} spread of  0.27.
Integration of our luminosity function gives a local volume density of AGN above $10^{41}$ erg s$^{-1}$
of $2.4\times 10^{-3}$ Mpc$^{-3}$, which is
about 10\% of the total luminous local galaxy density above $M_*=-19.75$. 
We have obtained X-ray spectra from the literature and from {\it Swift}  XRT follow-up observations.
These show that the distribution of $\log\: n_H$   is essentially flat
from $n_H=10^{20}$ cm$^{-2}$ to $10^{24}$ cm$^{-2}$,
with 50\% of the objects having column densities of less than $10^{22}$ cm$^{-2}$.
BAT Seyfert galaxies have a median redshift of 0.03, a maximum $\log$
luminosity of 45.1, and approximately half have
$\log n_H > 22$.
\end{abstract}


 \section{Introduction}

It is now realized that most of the AGN in the Universe
have  high column densities of absorbing material in our line of sight,
which significantly changes their  apparent  properties across much of the electromagnetic
spectrum. In many well studied objects this material
significantly reduces the soft X-ray, optical, and
UV signatures of an active nucleus essentially
``hiding'' the object.  
While it is commonly believed that extinction-corrected [OIII] 
can be used as an ``unbiased'' tracer of AGN
activity (Risaliti et al. 1999), there is a large 
scatter between [OIII] and $2-10$ keV X-ray flux 
(Heckman et
al. 2005) and between [OIII] and BAT flux 
(Mel\'endez et al 2008).  We acknowledge that some 
Compton thick AGN
are detected in [OIII] that cannot be detected in 
hard X-rays, but Compton thick AGN are outside the scope of
this paper.
Therefore,  surveys of AGN which
rely primarily on rest frame optical and UV studies are
 very incomplete and have led to misleading results
concerning the number, luminosity function, and evolution
of active galaxies  (e.g.,                       Barger et al. 2005).

While the distribution of column densities is under intensive
 investigation, it is clear from both X-ray (Tozzi et al. 2006,
 Cappi et al. 2006) and IR data (Alonso-Herrero, et al. 2006) that  a
large fraction of AGN  have column densities greater than
$3\times 10^{22}$ cm$^{-2}$ in the line of sight. Using the galactic
reddening  law (Predehl \& Schmitt 1995), this is equivalent to
$A_V>13$, making the nuclei essentially invisible in
the optical and UV bands. This effect seems to dominate
 the population seen in  deep X-ray
surveys (e.g., Barger et al. 2005, Brandt \& Hasinger 2005) where
 a large fraction of the X-ray selected objects do not have
 optical counterparts with classical AGN signatures.

There are only two spectral bands in which the nuclear emission is strong 
and  where, provided the column densities are less than 
 $1.5\times10^{24}$ cm$^{-2}$ (Compton-thin objects),  this obscuring material
is relatively optically thin. These bands, the hard X-ray ($E>20$ keV) and
the IR ($5-50\mu$m),  are optimal for unbiased searches for AGN (Treister et al. 2005).
 While recent results from {\it Spitzer} are finding many AGN via their IR emission, IR
 selection is  hampered by several effects (Barmby et al. 2006, Weedman et al. 2006,
Franceschini et al. 2006):
 (1) the strong emission from star
formation,
 (2) the lack of a unique ``IR color'' to  distinguish AGN
from other luminous objects (Stern et al. 2005),
   and
 (3) the wide range in IR spectral parameters (Weedman et al. 2006).
Thus,   while an IR  survey   yields  many objects,  it is very difficult to quantify 
its completeness and  how much of the IR luminosity of a particular galaxy is 
due to an  active nucleus. 
  These  complications are not present 
  in a hard X-ray survey  since at $E>20$ keV virtually
all the radiation comes from the nucleus and selection 
 effects are absent
  for Compton thin sources.
%
   Even for moderately Compton thick sources
 ($\Lambda < 2.3$ is absorption $< 90$\%),
        a hard X-ray survey has
   significant sensitivity, but without an 
   absorption correction the luminosity will be underestimated.
  Essentially every object more
 luminous that $10^{42}$ erg s$^{-1}$ is an AGN.  A hard X-ray survey is thus unique
in its ability to find all Compton thin AGN in a uniform, well-defined
fashion,  and to determine their intrinsic luminosity.
However, due to  the relative rarity of bright AGN    (even the {\it ROSAT}  all
sky survey has only $\sim$1 src deg$^{-2}$ at its threshold $-$ Voges et al. [1999]),
one needs  a very large solid angle survey to find the
 bright, easily studied objects.

 With the recent  {\it Chandra} and  {\it XMM} data
(e.g., Alexander et al. 2003, Giacconi et al. 2002, Yang et al. 2004;
       Mainieri et al. 2002, Szokoly et al. 2004, Zheng et al. 2004,
   Mainieri et al. 2005, Barger et al. 2001, 2003)
there has been great progress in understanding the
origin of the X-ray background and the evolution of AGN. It is now clear
that much of the background at $E>8$ keV is not produced
by the sources detected in the $2-8$ keV band
(Worsley et al. 2005), and is likely to  come from
a largely unobserved population of AGN with high column density and  low
redshift $z<1$.   Thus the source of the bulk of the surface
brightness of the X-ray background,
which peaks at $E\sim$30 keV (Gruber et al. 1999) is  uncertain.
The measurement of the space density and
evolution of this putative population of highly absorbed
AGN and the derivation of the distribution of their column densities
as a function of  luminosity and of redshift
 is crucial for modeling the X-ray background and
the evolution of active galaxies. Progress in this
area requires both a hard X-ray survey of sufficient sensitivity,
angular resolution and solid angle coverage to find and
identify large numbers of sources, {\it  and} follow-up observations with softer
X-ray measurements to obtain precise positions and
detailed X-ray spectral properties.

Due to a lack of instrumentation  with sufficient angular
 resolution to permit  identification of unique counterparts in other
wavelength bands and with sufficient solid angle and
sensitivity (Krivonos et al. 2005)  to produce a large
sample, there has been
little progress in hard X-ray surveys  for over 25 years
(e.g. Sazonov et al. 2005, 2007). 
   This situation has been radically changed by the
{\it Swift} BAT survey
(Markwardt et al. 2005)  and recent {\it INTEGRAL} results
(Beckmann et al. 2006b, Sazonov et al. 2007,
 Krivonos et al. 2005,  Bird et al. 2007) which have
detected more than 100 hard X-ray selected AGN, thus
providing the first unbiased sample of Compton thin
       AGN in the local Universe.
%

 \begin{figure}
 \centering
 \epsscale{1.15}
 \plotone{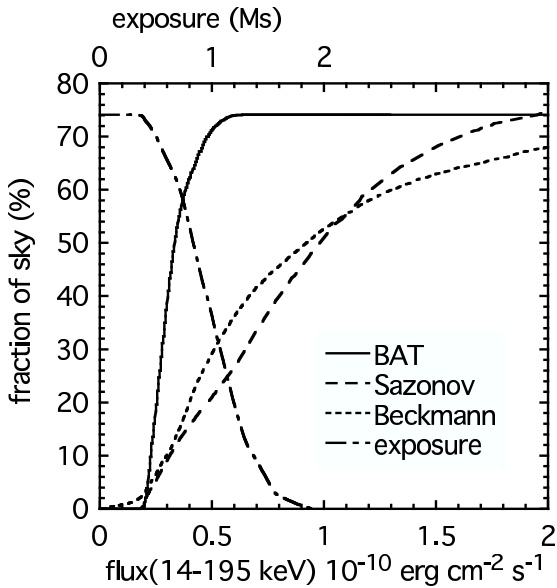}
 \vskip -2.cm
 \figcaption{
Percentage of the sky covered as a function of limiting 
flux in erg cm$^{-2}$s$^{-1}$ (14$-$195 keV) and of 
effective exposure (upper scale). As only the sky 
$|b|>15\degr$ is considered here, the maximum value is 
74\%. The corresponding curves as a function of limiting flux for 
the analyses of {\it INTEGRAL}  
data by Beckmann et al. (2006b) and by Sazonov et al. (2007)
 are shown for comparison, the flux having been 
converted assuming a power law spectrum with index $-2$.
 \label{sky_cov_plot}}
 \smallskip
 \end{figure}


In this paper we describe results from the first
9 months of the hard X-ray survey using the BAT instrument
 (Barthelmy et al. 2005) on the {\it Swift} mission
(Gehrels et al. 2005), concentrating on
 sources with $|b|> 15\degr$.  Above this latitude limit, we have identified 
all but  one of the sources detected at  $>4.8\sigma$  with optical
counterparts using {\it Swift} XRT and archival X-ray data.
With these same data we have also obtained   X-ray spectra.
 With a median positional uncertainty of 1.7' and a sensitivity limit
of a few times $10^{-11}$ erg cm$^{-2}$ s$^{-1}$ in the $14-195$ keV band,
the BAT data are about 10 times more sensitive 
than the previous all-sky hard X-ray
survey ({\it HEAO 1} A-4: 
   Levine et al. 1984) and the positions are accurate
enough to  allow unique identifications of nearly all of the sources.

Spectra are characterized by a photon index $\Gamma$, where 
$N(E)\propto E^{-\Gamma}$. Luminosities are calculated using $h_{70}=1, \:\Omega=0.3$.

\section{BAT Survey}

The second BAT catalog is based on the first 9 months of BAT
data (starting mid December 2005) and has several refinements compared
to the catalog of the first 3 months of data (Markwardt et al. 2005). The combination
of increased exposure, more uniform sky coverage and improved
software has increased  the total number of BAT sources by a factor $\sim$2.5.

\begin{figure}
 \centering
 \plotone{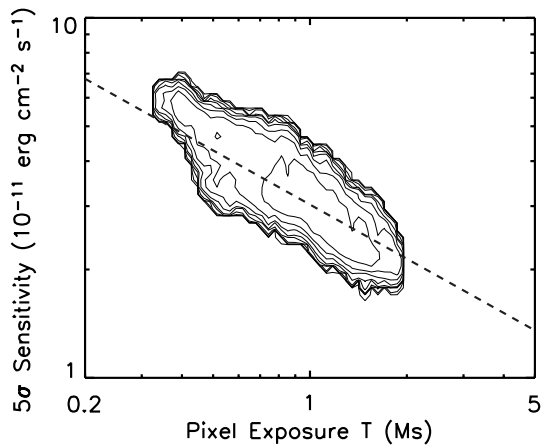}
 \vskip .75cm
 \figcaption{
  BAT survey $5\sigma$ sensitivity in the 
$14-195$ keV band for $|b| > 15\degr$ as a function
  of exposure.  The contours, spaced at logarithmic intervals,
 indicate the number of pixels ($|b|>15\degr$) 
in the all-sky mosaic with a given exposure and sensitivity.
  The dashed line indicates the survey sensitivity curve of Markwardt
  et al. (2005), without adjustment.
 \label{sens_plot}}
 \smallskip
 \end{figure}

We show the sky coverage in Figure \ref{sky_cov_plot}  and the sensitivity of the
survey as a function of exposure in Figure \ref{sens_plot}.  
There is a loss of sensitivity due to increased noise at low galactic 
latitudes from nearby bright sources,  and because of spacecraft constraints there tends to be somewhat reduced exposure in directions close to the ecliptic plane. Nevertheless the sensitivity achieved is comparatively uniform.   

\begin{figure}
 \centering
 \epsscale{1.15}
 \plotone{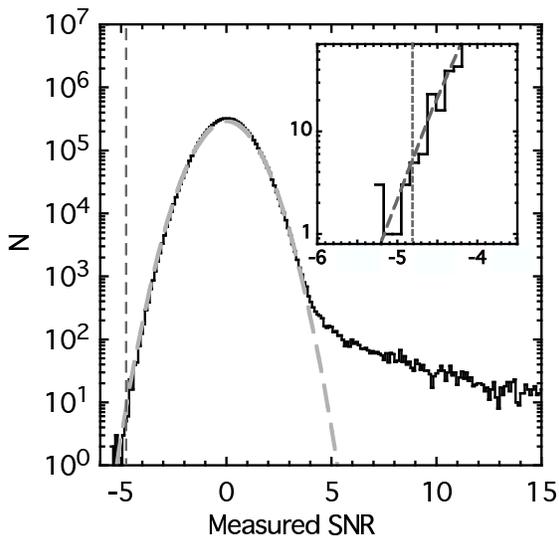}
 \vskip -2.cm
 \figcaption{
Histogram of the pixel values at $|b|>15\degr$  
in the 9 month survey all sky map relative to the 
local estimated noise level. 
 The data closely follow a Gaussian 
distribution with $\sigma=1.024$  except for the tail 
at high  positive values due to sources.  
The insert shows an expansion of the region 
below $SNR=-4$. Because of oversampling, 
 more than one pixel corresponds to  a single source.
\label{labelx}}
 \smallskip
 \end{figure}

We have picked a significance threshold of
$4.8\sigma$, which, based on the distribution of negative pixel residuals
(Figure \ref{labelx}), corresponds to a probability of $\sim$1  false source in the catalog.
In Table 1 we show all the sources detected at 
 $>4.8\sigma$ and with $|b|>15\degr$. The table also includes  
     sources that have been confidently identified 
with AGN but that lie at $|b|<15\degr$ or, while having significances 
less than $4.8\sigma$ in the final analysis have appeared 
at higher significance in partial or preliminary analyses.
Of the 44 AGN presented in Table 1 of Markwardt et al. (2005), only J1306.8$-$4023 does not appear in Table 1
of this study.
The spectral type is from V\'eron-Cetty \& V\'eron (2006), and where that is not available, 
we examined 6DF, SSDS or our own
observations and classified the AGN.
There are seven objects that do not have an 
optical classification, of which 2 have not been observed and the
remainder do not have optical AGN lines.

\begin{figure}
 \centering
 \plotone{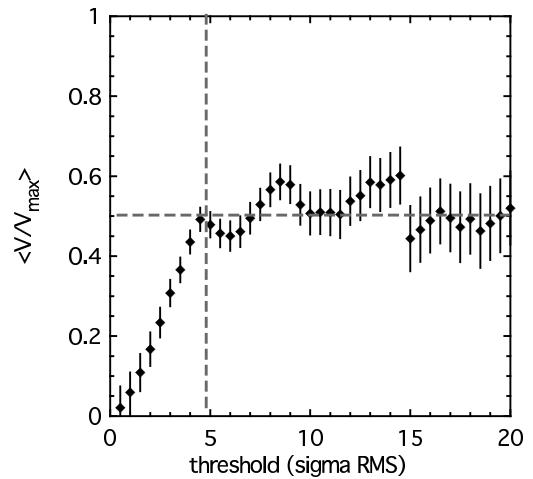}
 \vskip -1.25cm
 \figcaption{
Plot of $<V/V_{\rm max}>$ as a function of the significance
  threshold $\sigma$. For $\sigma>4.5$ the average ratio is consistent
with the nominal  $<V/V_{\rm max}>$ value of $0.5$.
 \label{vovmax_plot}}
 \end{figure}

We have verified the completeness of our sample by examining
the values of  $V/V_{\rm max}$ as a function of significance.  Above 
$4.8\sigma$ detection significance  we find a value of $0.5$, 
as expected for a complete sample from a uniform distribution (Figure \ref{vovmax_plot}).

\begin{figure}
 \centering
 \plotone{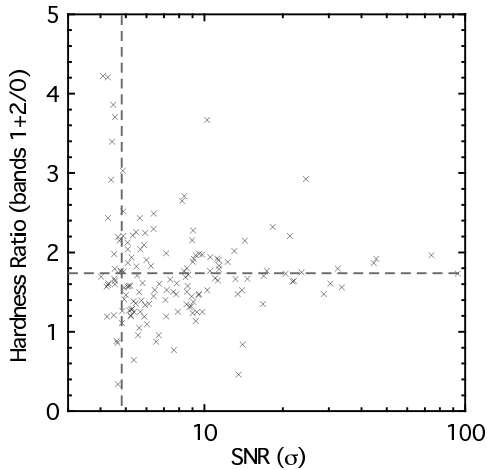}
 \vskip -1.cm
 \figcaption{
 Hardness ratio (Counts [$25-100$ keV]/Counts [$14-25$ keV]) 
as a function of  detection significance. There is no 
indication of discrimination against sources with 
soft spectra  near the 4.8$\sigma$ survey threshold.
\label{labelz}}
 \smallskip
 \smallskip
 \end{figure}

Basing the detection on significance in the total $14-195$ keV band is close to optimal
for sources with average spectra. We might miss some sources because their
spectra are much steeper. However,  as shown in Figure \ref{labelz}, there is no apparent
correlation between BAT hardness ratio and detection significance and thus we
believe that this selection effect is negligible in the present sample.

Because source detection is based on the entire 9 months of
data, it is possible that some sources might have been missed
if they had been very bright for only a fraction of the
observing time. This is confirmed by comparing the present
results with those of Markwardt et al. (2005). We found
that 9 of the Markwardt et al. sources do not lie above our
significance threshold of $4.8\sigma$ in the 9 months data.

\begin{figure}
 \centering
 \plotone{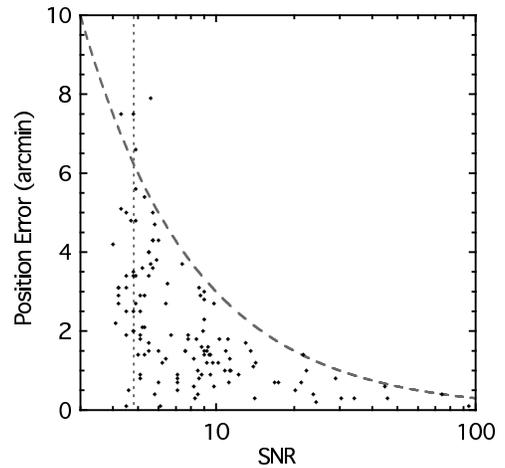}
 \vskip -1.cm
 \figcaption{
The distribution of mean offsets between positions measured 
with BAT and the counterpart as a function of the detection 
significance, {\it SNR}.  The     dashed line  corresponds  
to $30/${\it SNR}, or 6 arcmin at 5$\sigma$ significance.
The vertical dotted line is at the $4.8\sigma$ threshold used
 in this study. Sources below this threshold are not
complete and have been identified because their known spectrum 
is consistent with the BAT result.  Note that near the
threshold the errors can occasionally be larger than 
this model predicts.
\label{pos_err_plot}}
 \smallskip
 \smallskip
 \end{figure}

The accuracy of source positions (Figure \ref{pos_err_plot}) based on the
total AGN sample, depends  on significance, however,
at the significance limit of  $4.8\sigma$  of our
survey, the maximum 2$\sigma$ error circle radius is $\sim$6'.

\section{Sample Identification}

BAT is a wide field ($\sim2$ steradians) coded aperture 
hard X-ray instrument (Barthelmy et al 2006). During normal
operations it usually covers $\sim60$\% of the sky each day 
at $<20$ milliCrab sensitivity.  The BAT spectra were derived from
an all sky mosaic map in each energy bin averaged over 9 months 
of data beginning on 5 Dec 2004. The survey was
processed using the BAT Ftools\footnote{\tt http://heasarc.nasa.gov/ftools/ftools\_menu.html}  
                                             and additional 
software normalize the rates to on axis and to make mosaic
maps. The intrinsic binning in the BAT survey data product 
has 80 energy bins but to reduce processing time we used 4
energy bins for this survey. The energy bin edges are 
14, 24, 50, 100, 195 keV for the 9 month survey, but will be
expanded to 8 bins in the 22 month survey by dividing each 
of the current bins. The energies are calibrated in-flight
for each detector using an on-board electronic pulser and 
the 59.5 keV gamma-ray line and lanthanum L and M K X-ray
lines from a tagged $^{241}$Am source. The average count 
rate in the map bin that contains the known position of the
counterpart was used. Due to the the strong correlation of 
the signal in adjacent map bins of the oversampled coded
aperture image, it is not necessary to perform a fit to 
the PSF.  Each rate was normalized to the Crab nebula rate
using an assumed spectra of $10.4E^{-2.15}$ 
ph cm$^{-2}$ s$^{-1}$ keV$^{-1}$ for the BAT energy 
range. Due to the large number of
different pointings that contribute to any position in the 
map, this is a good approximation of the average response.
This has been verified by fitting sources known to have 
low variability and generally produces a good connection to
X-ray spectra in sources.  Error estimates were derived 
directly from the mosaic images using the RMS image noise in a
region around the source of roughly $3\degr$ in radius.  
This is the optimum procedure due to the residual systematic
errors of 1.2 to 1.8 times statistical values in the 
current BAT mosaics.  Analysis of the noise in the images
suggests that the variations in noise are small on this scale. 
Analysis
 of negative fluctuations shows that the noise is very well 
fit by a Gaussian distribution and that this
normalization is very accurate on average.  All fitting of 
the BAT data was performed on this normalized data using a
diagonal instrument response matrix. This procedure correctly 
accounts for instrumental systematics in sources with
spectral indices similar to the Crab. While there may be 
significant systematic errors for sources with spectra that
are much flatter than the Crab, this is not a significant 
problem for any of the sources presented in this paper.

\begin{figure}
 \centering
 \plotone{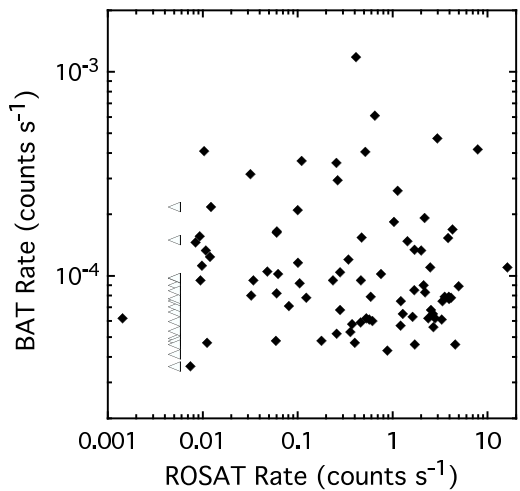}
 \vskip -1.cm
 \figcaption{
Comparison of {\it ROSAT} and BAT 
fluxes. Triangles indicate upper limits.
\label{rosat_v_bat}}
 \smallskip
 \smallskip
 \end{figure}

%
 We first attempted to identify the BAT sources using archival 
X-ray, optical, and radio data. The typical high
galactic latitude BAT source is a bright 
(2MASS $J$ band magnitude $> 13$) and nearby ($z<0.1$)
galaxy. While the counterpart is often a {\it ROSAT} or radio source, 
this is not a reliable indicator. In
particular we found little or no correlation between the BAT 
counting rates and the {\it ROSAT} all-sky survey
fluxes (Figure \ref{rosat_v_bat}), 
        making it difficult or impossible 
to utilize the {\it ROSAT} data to consistently identify the
sources. An examination of random positions suggests this type 
of source rarely falls in a BAT error circle.
While this approach was fruitful, we found a significant 
number of objects with either no obvious counterpart
or multiple possible counterparts, due to clustering.
We have followed up with {\it Swift} XRT
all but one of  the  BAT sources in the second catalog that did not have
 evident  identifications with previously known AGN, or
that  did not have   archival X-ray  measurements of  absorption column $n_H$
 from {\it XMM}, {\it ASCA}, {\it Chandra} or {\it Beppo-Sax}.
 We find that if the {\it Swift} XRT exposure is on the order of 10 ks or greater,
 we have a high probability of identifying an appropriate candidate.
 We define an appropriate candidate as one which is within  the 
 BAT $2\sigma$ error contour and whose X-ray flux is commensurate with the BAT detection.
Because of the possibility of source variability and of the low time resolution possible with the BAT
data ($\sim$2 weeks per significant data point)  
we require only that the X-ray flux is consistent with an 
absorbed power law model that has a flux  within a 
factor of ten of that predicted from the BAT detection.
A detailed analysis of the variability of the BAT data 
is presented in Beckmann et al (2006b) and a     comparison of the
XRT and other data in Winter et al (2008a).

We have  based our identifications on observations in the harder, $2-10$ keV,  part of the XRT band
to minimize the probability of a false identification.  A {\it Swift} XRT detection limit of
$0.001$ ct s$^{-1}$, or 10 total counts ($0.5-10$) keV in a 10 ks exposure,
corresponds to a $0.5-10$ keV flux of about 
$3.7\times 10^{-14}$ erg cm$^{-2}$ s$^{-1}$ 
for an unabsorbed source or to $6.3\times 10^{-14}$ 
erg cm$^{-2}$ s$^{-1}$ for one with an average $n_H$ of $10^{22}$.
Using the Moretti et al. (2003) $\log N-\log S$ distribution based on
{\it Chandra} data there are $\sim$50 or $20$ sources deg$^{-2}$, respectively, at these levels.
Thus the probability of finding a detectable source falling 
by chance within a $2\sigma$ BAT error circle  
($6$' radius at threshold) is high. 
However most of these sources would be expected 
to have a very low flux in the BAT band and thus not be
candidates for the counterparts of the BAT sources. 
We select the brightest source or sources at energies $>3$
keV as possible counterparts. A joint fit to the BAT and XRT 
data is performed using a simple spectral model
(partially covered power law) and allowing the relative 
normalization between the BAT and XRT data to be a
free parameter to account for variability. Agreement is 
defined as a relative normalization factor $<10$. A more
complex model is not usually required because the XRT 
data has insufficient statistical significance to constrain
complex models. See Winter et al (2008a) for a complete 
description. More complex models are required in a
few cases where our sources have very high column 
densities or are Compton thick (Winter et al. 2008c). 
These
cases are flagged in the table as complex. We have used 
similar criteria for identifications based on archival
data from other missions.

\begin{figure}
 \centering
 \epsscale{1.25}
 \plotone{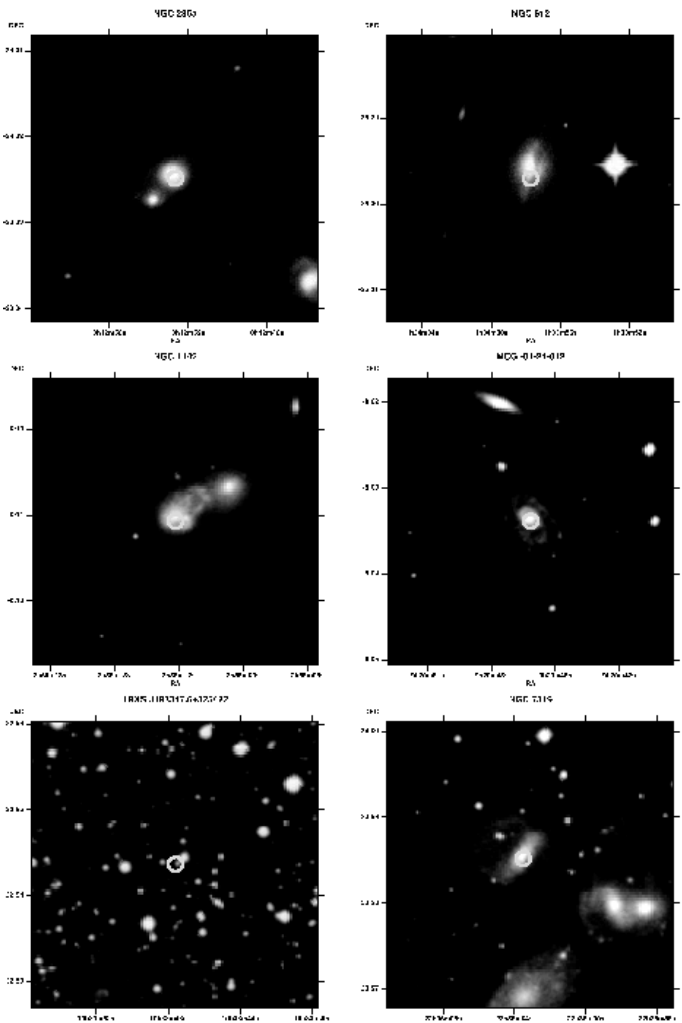}
 \figcaption{
Examples of the optical counterparts and the XRT error 
circles for sources detected with BAT.
\label{overlays}}
 \smallskip
 \end{figure}

When an XRT counterpart has been found, the error circle radius
is $\sim$4'',  and at the brightness of the optical counterparts (see below),
there is a very high probability of  identifying the object
in 2MASS or DSS imaging data. For all but one of the
      $|b|>15\degr$  sources there is a redshift
in the literature (based on NED), or from our follow-up program
 (Winter et al. 2008c) but often there is not
an available optical spectrum. Thus a significant number
of the objects do not have certain optical classifications.
We have used the optical spectral  types reported in V\'eron-Cetty \& V\'eron (2006) for AGN,
where available. In other cases we have used our own optical classifications based 
on SDSS or 6dF on-line data or what is available in NED and SIMBAD.
We show in Figure \ref{overlays} some of the optical counterparts and the XRT error circles.

With these criteria we have only one unidentified source 
 out of 130 sources with $\sigma> 4.8$ and $|b|>15\degr$, 
    but 13 out of 150 at $|b|<15\degr$.
 This difference arises from the much higher density of stars at lower galactic latitudes and to
the high degree of reddening and  lack of large spectroscopic surveys in the galactic
plane.    The relative completeness of the identifications in the BAT survey data  contrasts with that of
the {\it INTEGRAL} data  (Masetti et al. 2006a),   and Bird et al. 2007  and 
is due to the extensive XRT follow-up and the accurate
 positions possible with the XRT. The one unidentified  high latitude source 
 above $4.8\sigma$, SWIFT J1657.3+4807, has no reasonable X-ray counterpart
in the XRT field of view.   Obvious possibilities are (1) that this source is a  transient,
 or (2) that it  has  an extraordinarily high column density such
that the flux in the $2-10$ keV band is reduced by a
factor of $\sim$300,  e.g.,  a line of sight column density
of $>3\times 10^{24}$ cm$^{-2}$,    or a line of sight Compton optical depth of 2 (which
  would also require that there be no scattering into the line of sight greater than  0.2\%),
 or (3) that it is a ``false'' source, of which we expect $\sim 1$ in
the survey above our significance threshold.

We have examined the BAT light curves of all of the sources  in Table 1 (including those below
the $4.8\sigma$  threshold) and have determined that the
sources SWIFT J0201.9-4513,  SWIFT J0854.2+7221,
SWIFT J1319.7-3350, SWIFT J1328.4+6928 are almost certainly
transients.

\section{Results}

\subsection{Log N-Log S}

When investigating the  $\log N-\log S$ law, correct allowance for sky coverage 
near the detection threshold is crucial. The sky coverage as a function of 
limiting flux that we have used (Figure \ref{sky_cov_plot}) was obtained using
 the same measured RMS noise in the 9 month all-sky image that 
 was used in assessing source significances.
 This direct measure of sky coverage is much more reliable than
measures based on exposure as the systematic noise level
varies across the sky and is not a simple function of exposure.
At high fluxes the main uncertainties are due to 
Poisson statistics with a small number of objects.
At low fluxes they are associated with the correction 
for completeness, which is a strong function 
of the flux, that is itself uncertain.

\begin{figure}
 \centering
 \plotone{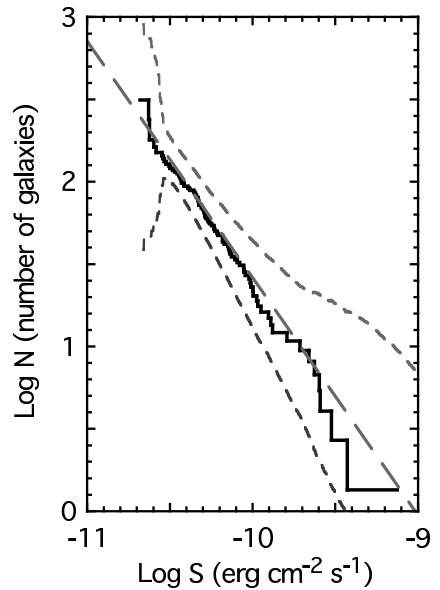}
 \vskip -2.cm
 \figcaption{
$Log\: N-Log\:S$ distribution for the BAT selected AGN.
 $S$ is in units of erg cm$^{-2}$ s$^{-1}$
 in the energy range 14$-$195 keV. 
The short-dashed lines show the 99\% confidence contours 
observed in Monte-Carlo simulations of observations of 
sources with a constant space density and 
the long-dashed lines a slope of $-1.5$.
  The long-dashed line is derived 
from the best fit to the differential spectrum in Figure \ref{logn_logs_diff}.
 \label{logn_logs}}
 \smallskip
 \end{figure}

\begin{figure}
 \centering
 \plotone{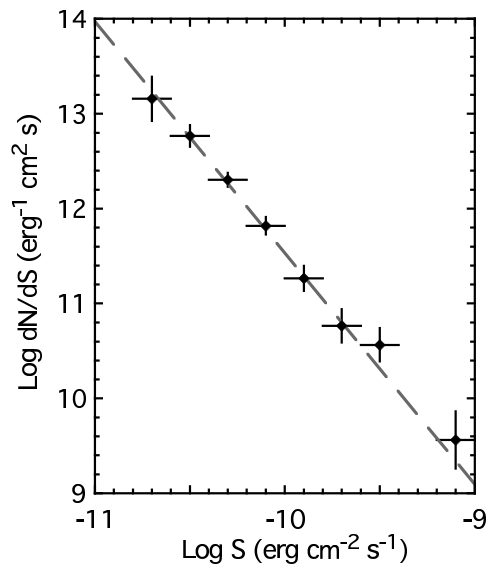}
 \vskip -1.cm
 \figcaption{
The differential $Log\: N-Log\:S$ distribution corresponding to 
 Figure \ref{logn_logs}. The fitted line has a slope of $-2.44\pm0.14$.
 \label{logn_logs_diff}}
 \smallskip
 \smallskip
 \end{figure}

 The $\log N-\log S$ distribution (Figures \ref{logn_logs},
                                  \ref{logn_logs_diff}) 
                    is well fit by the standard
  $S^{-3/2}$ function for uniformly distributed
sources and a normalization of $142.63\pm9.864$ AGN with flux $>3\times10^{-11}$ erg  cm$^{-2}$ s$^{-1}$.
  Formally we find a slope of $1.42\pm0.14$. Using a spectral
slope for each object, we can compare this $\log N-\log S$ law with those
derived from {\it INTEGRAL}  data (Beckmann et al. 2006b, Krivinos et al. 2005,
Sazonov et al. 2007). Converting our $\log N-\log S$ into
the Krivinos et al. $17-60$ keV band we find a normalization
which is $\sim$70\% of their value. Conversion into the $20-40$ keV band leads to 50\%
of the Beckmann et al. value. The most likely explanation of these differences
lies in the conversion factors used to convert
BAT or {\it INTEGRAL} counts to erg s$^{-1}$  (i.e., the instrument calibrations).
The Crab spectrum used by the Krivonos
et al. group for {\it INTEGRAL} calibration is $10\times E^{-2.1}$ (see Churazov et al. 2007
for a detailed discussion of the use of the Crab nebula
as a calibrator). The BAT team uses $10\times E^{-2.15}$.
In the $20-60$ keV band the {\it INTEGRAL} normalization
gives a Crab flux which is 1.15 higher. This would account for a
normalization of the $\log N-\log S$ law higher by a factor $1.23$,
very close to what is seen, and  consistent within the uncertainties.
The closeness of the BAT sample introduces some 
uncertainty in the distance measurement due to the random
velocities of galaxies ($\sim500$ km s$^{-1}$). To evaluate
 the effect of this uncertainty we have performed a Monte
Carlo simulation of the luminosity function, including the 
uncertainty in luminosity and in distance due to
the velocity error. This analysis indicates that the effect 
on the fitted parameters is $<1\sigma$. The break
log luminosity could be 0.2 dex higher due to this error 
compared with an noise error of 0.4. The largest effect
was on the high luminosity slope which could be 0.3 larger
  due to systematics (error 0.35). These systematic
  errors do not substantially effect the {\it Swift}/BAT 
  luminosity function at its current statistical accuracy.
Thus the $\log N - \log S$ law in the $14-195$ keV band is now 
 established to $\sim$25\% accuracy $-$ we know the number
 of sources quite accurately, but we  do not know their flux to better than 15\%.

\subsection{Luminosity Function}

\begin{figure}
 \centering
 \plotone{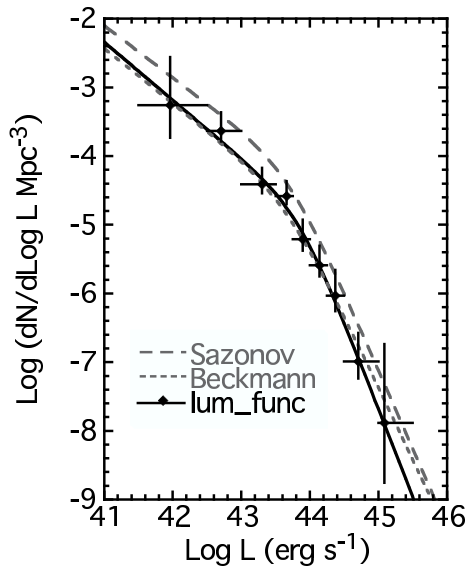}
 \vskip -3.cm
 \figcaption{
Comparison of the 14$-$195 keV  luminosity function derived from the BAT
observations with those found by Beckmann (2006b) 
and by Sazonov et al. (2007) using {\it INTEGRAL}. 
The {\it INTEGRAL} luminosities have been converted 
to the BAT band assuming an power law with photon index of 2.0.
 \label{lum_func_plot}}
 \smallskip
 \end{figure}

The high identification completeness of our survey and the good understanding of the sky coverage 
are important in finding the luminosity function. We use the standard broken power
law form
\begin{equation}
\Phi(L_X)= {A\over {\left[\left(L_X \over {L_*}\right)^a + \left(L_X \over {L_*}\right)^b\right] }},
\end{equation}
This provides an excellent
description of the data with the parameters given in Table \ref{lum_fun_tab}. 
For comparison of other observations with ours we have converted luminosities  quoted in other energy bands 
assuming a spectrum breaking from a 
slope of 1.7 to a slope of 2.0 at 10 keV. 
The BAT luminosity function shown in Figure \ref{lum_func_plot} agrees
well with those
 obtained by  
Beckman et al. (2006b) 
  and by Sazonov et al. (2007) using data from {\it INTEGRAL}
    both in terms of the slopes and the break luminosities,
    though their errors are 
generally somewhat larger.  However we find  a significantly lower break luminosity
than  found by  Barger et al. (2005) and by La Franca et al. (2005)  from observations at lower energies. 
The rather large difference cannot be caused by
spectral conversion factors that neglect absorption in the $2-10$ keV band,  since this would make the 
observed $2-10$ keV luminosity even lower  compared to the $14-195$ keV value, exacerbating the problem.
 We thus believe that the disagreement between the luminosity functions is due to a  
 deficit of objects at $\log L$(erg s$^{-1}$)$ < 44.11$  in the $2-10$ keV band. 
   Considering that the bulk of the objects and their emitted 
  luminosity lies near the break luminosity, this  could imply a substantial modification to the 
  present day evolution models (e.g., Gilli, Comastri \& Hasinger 2007).

As we show in the next section, the probability of an object
being absorbed is a  function of $14-195$ keV luminosity.
Hence  there is a strong selection against detecting low
luminosity AGN in softer X-ray surveys (see the discussion
in Sazonov et al. 2007).

\subsection{Nature of the Identifications}

There are 151 sources in Table 1 which we have identified 
with AGN. 102 are at high latitude ($|b|>15\degr$) and above
   4.8$\sigma$ and form our complete sample.  The remainder 
 are at low latitude (42) and/or have lower significance 
in the final analysis (44).  In the complete sample 14 out
 of 102 are beamed sources -- BL Lacs and Blazars -- 
(17 out of 152 overall) and the remainder are Seyferts and 
galaxies which show indications of activity.  In addition,
 we have detected 32 galactic sources and 2 galaxy clusters 
which meet the  latitude and significance criteria for 
the complete sample.  At low latitudes we also detect at 
$>4.8\sigma$  103 galactic sources, 3 galaxy clusters, 
and 13 unidentified sources.
%
%
Although they are included in Table 1,  we have not used sources 
identified as blazar or BL Lac,  nor any source with
$z>0.5$, in the distribution functions. 
%

 We use the $J$ band magnitudes
from the 2MASS survey to categorize the objects since that
is the largest homogeneous data base which covers the largest
fraction of the {\it Swift} BAT sources. It is noticeable that    
the faintest optical counterparts are the blazars and
the galactic sources. The optically determined AGN tend to be
in fairly bright galaxies.      One of the reasons that there
are so few blazar identifications at low galactic latitudes
is the relative faintness of the likely optical counterparts
combined with the lack of available redshifts and the effect of galactic reddening.

 \begin{figure}
 \centering
 \plotone{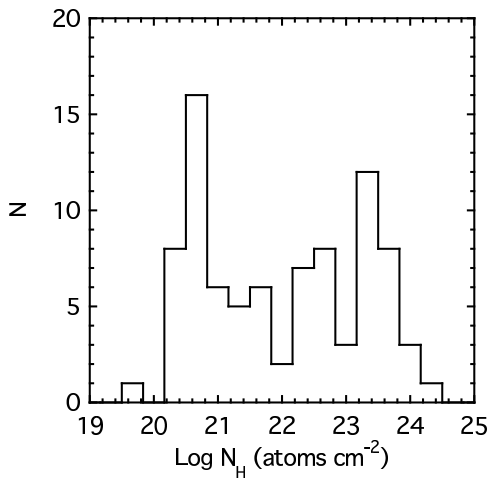}
 \vskip -2.cm
 \figcaption{
The distribution of column densities for the BAT selected AGN. Notice the peak at low 
column densities and the relatively flat distribution above it. 
The galactic column density has not been subtracted.
 \label{nh_distrib}}
 \smallskip
 \end{figure}

 \begin{figure}
 \centering
 \plotone{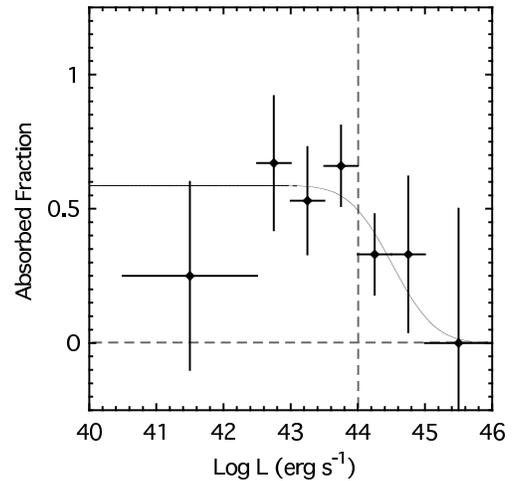}
 \vskip -1.25cm
 \figcaption{
The fraction of BAT selected AGN with $n_H>10^{22}$ cm$^{2}$ 
as a function of  14$-$195 keV luminosity. The  position of the 
break in the luminosity function slope is indicated. 
The smooth curve is simply one form which is consistent 
with the data. As elsewhere, only AGN with $|b|>15\degr$ and 
significance greater than 4.8$\sigma$ have been included. 
We note that if AGN with  $|b|<15\degr$ are included the drop at 
high luminosity is less pronounced but it is still 
significant at the $>2\sigma$ level. 
 \label{abs_frac}}
 \smallskip
 \smallskip
 \end{figure}

Nine  of the objects have not previously been optically
classified as AGN. An excellent example of this is the object
 NGC 4138 (Ho 1999, Moustakas \& Kennicutt 2006) which
shows little or no [OIII] emission and  in which only very
high signal to noise spectra revealed a very faint broad H$\alpha$
line. Other objects, like NGC4102 (Moustakas \& Kennicutt 2006)
 show no optical evidence of AGN activity.

For those objects which are optically classified as AGNs, 33 are
Seyfert 1s, 14 are Seyfert 1.5, 35 are Seyfert 2s.  There is
reasonable but not perfect correlation between the optical
classification and the presence of X-ray absorption (see below).
Only two  of 33     Seyfert 1's have   a     column density
greater than $10^{22}$ cm$^{-2}$, whereas    4 of 14      Seyfert 1.5's and
 33 of 35     Seyfert 2's are absorbed  (two do not have X-ray column densities).

The median redshift of the non-blazars is $\sim0.017$. However, the
 blazar redshift distribution is very different with a long
 tail to high redshift and a median redshift of 0.24 (mean of 0.76). Thus we have 
been careful in determining the overall luminosity function
to separate the blazars from the non-blazars since this will
significantly change the slope of the
high luminosity end of the luminosity function.

 \begin{figure}
 \centering
 \plotone{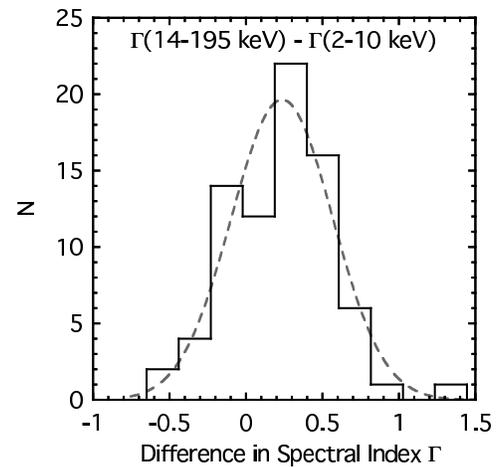}
 \vskip -1.95cm
 \figcaption{
 Histogram  of the X-ray spectral index in the 
BAT band minus the X-ray spectral index.  The X-ray 
indices are mostly from {\it ASCA} and XRT with some 
from various other missions. 
The mean difference is 0.26 with a
 standard deviation of  0.36.
 \label{slope_diff_plot}}
 \smallskip
 \smallskip
 \end{figure}

\subsection{X-ray Spectral Analysis}

The X-ray spectra of many of the sources have been published
(see the references in Table 1). In these cases we have used the 
previously reported values of the column densities of the sources, 
while noting that the signal to noise of the observations
varies greatly, as does the sophistication of the analysis and
the type of models used to classify the spectra.  Many of the
spectra are rather complex (Winter et al. 2008a), making
assignment of errors to the column density difficult and
highly model dependent.  Where the column densities
in Table 1 were  obtained with {\it Swift} XRT follow-up 
observations,  for homogeneity we report  the results of simple absorbed power law fits.
As shown in Figure \ref{rosat_v_bat}, a large fraction 
of the BAT sources are not detected 
by the {\it ROSAT} all sky survey, despite its factor of 100 better sensitivity for
unabsorbed sources. This graphically illustrates the
importance of obscuration in the selection of X-ray samples.

A detailed analysis of the archival
{\it XMM}, {\it ASCA}, {\it BeppoSax}, and {\it Chandra}
data as well as the {\it Swift}
XRT data will presented in another paper (Winter et al. 2008a).

 \begin{figure}
 \centering
 \plotone{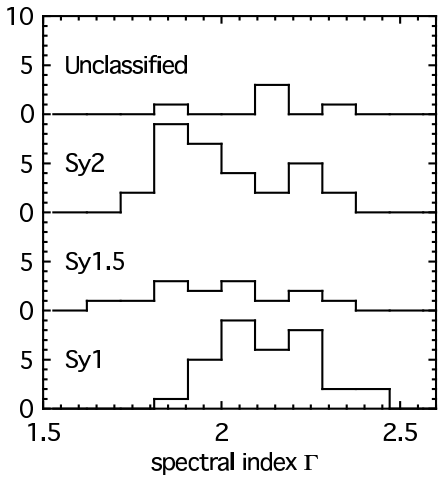}
 \vskip -2cm
 \figcaption{
Distribution of power law indices in the $14-195$ keV band for BAT selected sources 
sorted into Seyfert 1, 
Seyfert 1.5, Seyfert 2 and unclassified objects.
 \label{pl_index_distrib}}
 \smallskip
 \end{figure}

The distribution of absorption  for the non-blazars
  (Figure \ref{nh_distrib}) 
             is almost flat for $ \log n_H$(cm$^{-2}$) in the range $21-24$,  
    with a strong peak at low column density due primarily to the effects of galactic obscuration.
The relative paucity of  Compton thick objects ($ \log n_H$[cm$^{-2}$]$\ge24.5$)  is interesting.
 Unfortunately  at such high columns the flux, even in the BAT energy band, 
is severely reduced so our level of completeness is uncertain.  In addition we are only able to fit simplified models
 for many of these objects. Thus quantification of the lack of Compton thick objects 
 awaits  more observations with  high sensitivity X-ray spectrometers (e.g.,
 {\it XMM}, {\it Suzaku}).

As shown in Figure \ref{abs_frac},  
the fraction of strongly absorbed AGN drops with increasing luminosity.
This is consistent with the previous claims 
of a drop in the absorbed fraction at higher 
luminosities, but it is not yet
of sufficient statistical significance
to confirm this dependence.
While this has been seen in several X-ray selected surveys
(Ueda et al. 2003, La Franca et al. (2005),
Shinozaki et al. 2006), the fact that the selection of BAT
sources is independent of the line of sight column
density confirms and extends these results.

\subsection{BAT Spectral Analysis}

At the present stage of analysis we only have four channel
spectra available (this is a limitation  of the present analysis
software and is not intrinsic to the experiment). We have thus
fit only simple power law models to the data.

The fact that the  BAT hardness ratio shows no correlation with signal-to-noise 
  (Figure \ref{labelz}) indicates that there is no selection 
   bias due to spectral parameters.
The median spectral index is $\Gamma=1.98$,  in agreement 
with the {\it INTEGRAL} results from  Beckmann et al. (2006b), with an {\it rms} spread 0.27.
For a sample of 74 sources which have archival X-ray spectrum  spectra at lower energies
(e.g., Markowitz \& Edelson 2004),   the BAT slope
is on average $\sim0.23$ steeper than in the X-ray band 
(Figure \ref{slope_diff_plot}). 
A viable explanation for
this (Nandra et al. 1999) is that  the BAT data are detecting the
``true'' X-ray spectral slope of 2, while the X-ray data are
 strongly influenced by the effects of reflection.   
Malizia et al. (2003) found using  {\it BeppoSAX} hard X-ray 
data that Seyfert 2s are systematically harder than Seyfert 1s. 
A similar result is reported by Beckmann et al. (2006a).  
Comparison of the spectral index distributions of 
Seyfert 1 and Seyfert 2s (Figure \ref{pl_index_distrib}) 
confirms this finding --  according to a Kolmogorov-Smirnov test
the two distributions have a probability of less 
than $0.1\%$ of arising from the same parent 
 distribution function.

\section{Discussion}

\subsection{Luminosity Function}

As shown above the low luminosity slope of the 
{\gks luminosity function of} hard X-ray
selected AGN is steeper than that of the $2-8$ keV function
of Barger et al. 2005. We believe that this is  due to the
high fraction of heavily absorbed objects at low
BAT luminosities. Thus the contribution of low luminosity
 objects to the $10-100$ keV background is larger than
originally calculated.   This is confirmed by the agreement
of the slope of our luminosity function with the absorption
corrected low luminosity slope of La Franca et al. (2005),
which unlike Barger et al (2005)
assumes an absorption that depends on luminosity.
The break in the luminosity function is quite robust and
thus is an intrinsic feature of the luminosity function
  and is not due to a spectral selection effect.
Integration of our luminosity function gives
a local volume density of
$n(L_X>10^{41}$erg s$^{-1})= 2.4\times 10^{-3}$ Mpc$^{-3}$, compared to a
density of $0.02$ Mpc$^{-3}$  galaxies brighter than  $M_*=-19.75$ (Cross et al. 2001), 
and a local emissivity of 2.3$\times 10^{39}$ erg  s$^{-1}$ Mpc$^{-3}$.
The choice of  $M_*$ is that is the knee in 
the luminosity function and is the typical absolute magnitude for a galaxy.
It is a simple way of estimating the galaxy density.
%
The typical $J$ band absolute magnitude at the knee is
$M*=-21.73$ (Cole et al 2001). The median BAT $J$ 
band absolute magnitude is $M=-23.8$ and only 3 BAT AGN have
$M>-22$. 
Hence  $\ga10$\% of luminous galaxies in the local Universe are AGN
with a hard X-ray luminosity $\ga10^{41}$ erg s$^{-1}$.
Because of the low median redshift of the sample, the BAT data are not sensitive 
to evolution in the luminosity function and  $V/V_{\rm max} \sim$0.5  is as expected.

\subsection{Log N-Log S}

There have been numerous predictions of the
hard X-ray $\log N-\log S$ (Treister et al. 2006,
Gandhi \& Fabian 2003) and our data allow a direct comparison
of these models. We find that converting the
observed BAT $\log N-\log S$ to the band predicted by these
 authors that we have good agreement with the
predictions of Gandhi et al. (2004), but lie a factor of 2 lower than that
predicted by Treister et al. (2006). Since each of these models
makes different assumptions, our hard X-ray survey
should be able to {\gks determine which are valid.} 

\subsection{The distribution of $n_H$}

In 
   Figures \ref{abs_frac} and \ref{pl_index_distrib}
   the distribution of column 
densities over all objects is almost flat and appears to depend
on hard X-ray luminosity. Similar results based on the {\it RXTE} 
slew survey were obtained by Sazonov \& Revnivtsev (2004).
The standard unified model predicts that the ratio of 
absorbed to unabsorbed objects should be $4:1$,
  as opposed to our
observed value of $1:1$. This difference is probably due 
to the neglect of the luminosity dependence of absorption in
the simple unified model. The BAT results are roughly consistent 
with dependence of absorption on luminosity seen previously
(Ueda et al. 2003, Steffen et al. 2003, Gilli et al. 2007). 
We note that the distribution of column densities in Tozzi
et al. (2006) from the {\it Chandra} deep fields is rather 
different from the BAT sample in that the Tozzi et al. sample seems
to be missing the low $n_H$ half of the distribution. 
This has been confirmed by Wang et al. (2007) and by Gilli et al.
(2007).
Direct comparison of the $n_H$ distribution from 
the BAT sample and Tozzi et al shows apparent differences,
especially at low $n_H$. Taken at face value, this 
would indicate an evolution of the $n_H$ distribution between
the low median redshift of the BAT sample (0.03) and 
the redshift of the Tozzi sample ($\sim0.7$). This is
similar to the results reported by La Franca et al. (2005), however 
 Hasinger et al (2008) find no such dependence.

\section {Conclusion }

We have presented the results of an AGN survey
using data from the BAT instrument on {\it Swift}.
 The use of a hard X-ray bandpass means that the
survey is immune to the effects of X-ray absorption that
have traditionally plagued similar studies
 in optical and soft X-ray bandpasses,
 raising serious  questions concerning   completeness.
Utilizing the standard AGN broken power
law  prescription to characterize the differential luminosity
distribution function, we find that the data can be very well
described taking  a break luminosity
$\log L_*$(erg s$^{-1}$)$=43.85\pm0.26$,
  a  low luminosity power law slope
$a=0.84^{+0.16}_{-0.22}$, 
                  and a  high luminosity power law slope
  $b=2.55^{+0.43}_{-0.30}$,
                 in agreement with other studies based on
hard X-ray survey data such as that of Sazonov et al. (2007) using {\it INTEGRAL}.
We find a median spectral index $1.98$,
in accord with the Beckmann et al. (2006b) study
using {\it INTEGRAL}.
 By integrating  our inferred  luminosity function above
$10^{41}$ erg s$^{-1}$, we
arrive at a  local volume density of
$2.4\times 10^{-3}$ Mpc$^{-3}$,
roughly 10\% of the local density of luminous galaxies.

The BAT survey has detected 31 AGN at $>4.8\sigma$ that were 
not previously detected in hard X-rays, of which  9 were
not previously identified as AGN by other techniques.  In addition,
 there are 14 BAT AGN that were also detected
contemporaneously in hard X-rays by {\it INTEGRAL}, of which 
5 had not been previously identified as AGN. For sources that
were detected by both instruments, there is a good correlation 
between the BAT and {\it INTEGRAL} flux, with the exception
of a few sources that are almost certainly variable. There are 
42 {\it INTEGRAL} AGN with SNR $>4.8$ that were not detected by
BAT.  Only 11 of these have a flux (scaled to the BAT energy 
band assuming $E^{-2}$ spectrum) that is greater than 
$3\times10^{-11}$ erg cm$^{-2}$ s$^{-1}$, where a 
BAT detection is likely.   Most of these high-flux, 
undetected sources are within $30\degr$
 of the Galactic Center, where the BAT survey has 
significantly reduced sensitivity due to lower exposure and
increased systematic errors.  Of the BAT detected sources,
  13\% were not previously known to be AGN.

With increased exposure, both the BAT and {\it INTEGRAL} 
survey sensitivities will improve, and 
we expect most of the new
unidentified hard X-ray sources to be in the 
interesting class of very heavily absorbed AGN. 
{\it INTEGRAL} detected 111 AGN
at $>4.8\sigma$ in $\sim4$ yr. Due to its larger FOV 
and random observing strategy, BAT detected 126 AGN in $0.75$ yr, a
rate 6 times faster than {\it INTEGRAL}. We expect 
both missions to continue accumulating new AGN at the same rates, in
which case BAT AGN will become an increasing fraction of 
the new detections. At 3 yr after the {\it Swift} launch, we
predict 450 BAT detected AGN and more than 60 that not have
 been previously identified as AGN. The hard X-ray
measurements are unique in another sense. We believe they 
yield a accurate measurement of the average luminosity of
these sources.  We have shown (Winter et al. 2008bc) that 
the luminosity and power law index for absorbed sources cannot be
accurately derived from $2-10$ keV X-ray measurements alone, 
even with {\it XMM} or {\it Chandra}.  For the $\sim1/2$ of all 
AGN that are
absorbed, the BAT and {\it INTEGRAL} surveys provide a 
unique new measurements of the luminosity and underlying power law.

This is the second paper in a series. In future papers
we will present the X-ray spectral properties of these objects,
the long term  BAT light curves, detailed spectral analysis
of the BAT data  and  the optical properties of the hosts
of the BAT sources,  and extend the sample by a factor
of two in size.

\def\mnras{MNRAS}
\def\apj{ApJ}
\def\apjs{ApJS}
\def\apjl{ApJL}
\def\aj{AJ}
\def\araa{ARA\&A}
\def\aap{A\&A}

\clearpage

%
















\vfil\eject
\pagestyle{empty}
\setlength{\voffset}{20mm}

\LongTables
\clearpage
\begin{landscape}
\voffset=-.235truein
\begin{deluxetable}{ rllrrcrclccccccrr}
\small
\addtolength{\tabcolsep}{-5pt}
\tabletypesize{\tiny}
\tablewidth{0pt}
 \tablecaption{{\it Swift} Survey Table.} 
\tablehead{
\colhead{\hspace{21.5mm}\#} &
\colhead{\it Swift\tablenotemark{a}} & 
\colhead{ID\tablenotemark{b}} & 
\colhead{\hfill RA\tablenotemark{c} \hfill} & 
\colhead{\hfill Dec\tablenotemark{c} \hfill}  & 
\colhead{$>15\degr$} & 
\colhead{\rm SNR} & 
\colhead{\hspace{-4mm} $f_{\rm BAT}$} & 
\colhead{$\hfill z \hfill$} &  
\colhead{$\log L\:\:$\tablenotemark{e}} &
\colhead{$\log n_H$} & 
\colhead{Ref.} & 
\colhead{Cmplx} & 
\colhead{Type} & 
\colhead{Note} &  
\colhead{$J$ \hspace{0.5mm}} & 
\colhead{$f_{\rm ROSAT}$} \\ 
\colhead{\hspace{8mm}} &
\colhead{name} &
\colhead{ } & 
\colhead{\hfill deg \hfill} & 
\colhead{\hfill deg \hfill} & 
\colhead{\tablenotemark{d}} & 
\colhead{ } & 
\colhead{\tablenotemark{e}} &
\colhead{ }  &
\colhead{erg   s$^{-1}$} &
\colhead{cm$^{-2}$} &
\colhead{\tablenotemark{f}} & 
\colhead{\tablenotemark{g}} & 
\colhead{\tablenotemark{h}} &
\colhead{\tablenotemark{i}} &
\colhead{mag} & 
\colhead{rate\tablenotemark{j,k}}
}
\startdata
    1 & \hspace{2mm} SWIFT J0042.9$-$2332 \hspace{2mm} & NGC 235A & 10.7200 & $-$23.5410 & y~ & 4.47 & 3.2 \hspace{000mm} & 0.022229 & 43.56 & 23.00 & \ref{SwiftXRT} & y & Sy 2\dag & & 10.58 & 0.024 \\
    2 & \hspace{2mm} SWIFT J0048.8+3155$^l$ \hspace{2mm} & Mrk 348 & 12.1964 & 31.9570 & y* & 13.00 & 9.5 \hspace{000mm} & 0.015034 & 43.68 & 23.32 & \ref{Akylas02} & y & Sy 2 & & 11.24 & 0.009 \\
    3 & \hspace{2mm} SWIFT J0059.4+3150 \hspace{2mm} & Mrk 352 & 14.9720 & 31.8269 & y* & 4.90 & 3.7 \hspace{000mm} & 0.014864 & 43.27 & 20.75 & \ref{Winter_inprep} & & Sy 1 & & 12.49 & 0.615 \\
    4 & \hspace{2mm} SWIFT J0114.4$-$5522 \hspace{2mm} & NGC 454 & 18.5946 & $-$55.3986 & y~ & 4.54 & 2.3 \hspace{000mm} & 0.012125 & 42.88 & 22.95 & \ref{SwiftXRT} & y & Sy 2 & \ref{Veron_Veron01} & 13.98 & \\
    5 & \hspace{2mm} SWIFT J0123.9$-$5846$^l$  \hspace{2mm} & Fairall 9 & 20.9408 & $-$58.8057 & y* & 8.90 & 4.7 \hspace{000mm} & 0.04702 & 44.39 & 20.36 & \ref{Lutz04} & & Sy 1 & & 11.85 & 3.350 \\
    6 & \hspace{2mm} SWIFT J0123.8$-$3504$^l$  \hspace{2mm} & NGC 526A & 20.9766 & $-$35.0654 & y* & 8.20 & 5.2 \hspace{000mm} & 0.019097 & 43.63 & 22.30 & \ref{Lutz04} & y & Sy 1.5 & & 11.60 & 0.123 \\
    7 & \hspace{2mm} SWIFT J0134.1$-$3625 \hspace{2mm} & NGC 612 & 23.4906 & $-$36.4933 & y* & 4.89 & 3.2 \hspace{000mm} & 0.029771 & 43.81 & 23.70 & \ref{Sambruna98} & y & Gal/Radio & \ref{Lewis03} & 11.68 & \\
    8 & \hspace{2mm} SWIFT J0138.6$-$4001$^l$  \hspace{2mm} & ESO 297$-$018 & 24.6548 & $-$40.0114 & y* & 9.03 & 4.9 \hspace{000mm} & 0.025201 & 43.85 & 23.84 & \ref{SwiftXRT} & y & Sy 2 & & 9.18 & \\
    9 & \hspace{2mm} SWIFT J0201.0$-$0648 \hspace{2mm} & NGC 788 & 30.2769 & $-$6.8155 & y* & 8.37 & 5.9 \hspace{000mm} & 0.013603 & 43.39 & 23.48 & \ref{ASCA_RM_inprep} & y & Sy 2 & & 10.02 & \\
    10 & \hspace{2mm} SWIFT J0206.2$-$0019 \hspace{2mm} & Mrk 1018 & 31.5666 & $-$0.2914 & y* & 5.31 & 3.5 \hspace{000mm} & 0.04244 & 44.17 & 20.53 & \ref{SwiftXRT} & & Sy 1.5 & & 11.60 & 0.360 \\
    11 & \hspace{2mm} SWIFT J0209.7+5226 \hspace{2mm} & LEDA 138501 & 32.3929 & 52.4425 & & 5.13 & 3.9 \hspace{000mm} & 0.0492 & 44.34 & 21.18 & \ref{SwiftXRT} & & Sy 1 & & & 0.752 \\
    12 & \hspace{2mm} SWIFT J0214.6$-$0049 \hspace{2mm} & Mrk 590 & 33.6398 & $-$0.7667 & y* & 5.67 & 3.7 \hspace{000mm} & 0.02638 & 43.77 & 20.43 & \ref{Gallo06} & & Sy 1.2 & & 10.71 & 2.689 \\
    13 & \hspace{2mm} SWIFT J0216.3+5128 \hspace{2mm} & 2MASX J02162987+5126246 & 34.1243 & 51.4402 & & 4.93 & 3.6 \hspace{000mm} & & & 22.25 & \ref{SwiftXRT} & & Galaxy\dag & \ref{not_obs} & 14.27 & \\
    14 & \hspace{2mm} SWIFT J0218.0+7348 \hspace{2mm} & [HB89] 0212+735 & 34.3784 & 73.8257 & & 4.27 & 2.6 \hspace{000mm} & 2.367 & 48.05 & 23.38 & \ref{SwiftXRT} & & BL Lac & & & 0.044 \\
    15 & \hspace{2mm} SWIFT J0228.1+3118 \hspace{2mm} & NGC 931 & 37.0603 & 31.3117 & y* & 8.56 & 7.3 \hspace{000mm} & 0.016652 & 43.66 & 21.65 & \ref{Tartarus} & & Sy 1.5 & & 10.40 & 0.342 \\
    16 & \hspace{2mm} SWIFT J0234.6$-$0848 \hspace{2mm} & NGC 985 & 38.6574 & $-$8.7876 & y* & 5.07 & 3.7 \hspace{000mm} & 0.043 & 44.21 & 21.59 & \ref{Tartarus} & y & Sy 1\dag & & 11.63 & 1.281 \\
    17 & \hspace{2mm} SWIFT J0235.3$-$2934 \hspace{2mm} & ESO 416$-$G002 & 38.8058 & $-$29.6047 & y~ & 4.76 & 3.2 \hspace{000mm} & 0.059198 & 44.42 & $<19.60$ & \ref{XMM_RM_inprep} & & Sy 1.9 & & 12.15 & 0.356 \\
    18 & \hspace{2mm} SWIFT J0238.2$-$5213$^l$  \hspace{2mm} & ESO 198$-$024 & 39.5821 & $-$52.1923 & y* & 7.82 & 3.9 \hspace{000mm} & 0.0455 & 44.27 & 21.00 & \ref{Tartarus} & & Sy 1 & & 12.68 & 2.380 \\
    19 & \hspace{2mm} SWIFT J0244.8+6227 \hspace{2mm} & QSO B0241+622 & 41.2404 & 62.4685 & & 11.19 & 7.3 \hspace{000mm} & 0.044 & 44.52 & 21.98 & \ref{EXOSAT_RM_inprep} & & Sy 1 & & & 0.414 \\
    20 & \hspace{2mm} SWIFT J0255.2$-$0011$^l$  \hspace{2mm} & NGC 1142 & 43.8008 & $-$0.1836 & y* & 9.80 & 7.8 \hspace{000mm} & 0.028847 & 44.17 & 23.38 & \ref{XMM_RM_inprep} & y & Sy 2\dag & & 10.06 & 0.011 \\
    21 & \hspace{2mm} SWIFT J0318.7+6828 \hspace{2mm} & 2MASX J03181899+6829322 & 49.5791 & 68.4921 & & 4.89 & 3.5 \hspace{000mm} & 0.0901 & 44.85 & 22.59 & \ref{SwiftXRT} & & Sy 1.9 & \ref{Schoenmakers98} & 15.13 & \\
    22 & \hspace{2mm} SWIFT J0319.7+4132 \hspace{2mm} & NGC 1275 & 49.9507 & 41.5117 & & 13.51 & 11.5 \hspace{000mm} & 0.017559 & 43.90 & 21.18 & \ref{Bassani99} & & Sy 2 & & 11.02 & 4.756 \\
    23 & \hspace{2mm} SWIFT J0328.4$-$2846 \hspace{2mm} & PKS 0326$-$288 & 52.1521 & $-$28.6968 & y~ & 4.50 & 2.3 \hspace{000mm} & 0.108 & 44.84 & & & & Sy 1.9 & \ref{6dF} & 14.19 & \\
    24 & \hspace{2mm} SWIFT J0333.6$-$3607$^l$  \hspace{2mm} & NGC 1365 & 53.4015 & $-$36.1404 & y* & 13.93 & 7.2 \hspace{000mm} & 0.005457 & 42.67 & 23.60 & \ref{Lutz04} & y & Sy 1.8 & & 7.36 & 0.101 \\
    25 & \hspace{2mm} SWIFT J0342.0$-$2115 \hspace{2mm} & ESO 548$-$G081 & 55.5155 & $-$21.2444 & y* & 5.45 & 3.3 \hspace{000mm} & 0.01448 & 43.19 & 20.48 & \ref{SwiftXRT} & & Sy 1 & & 9.35 & 0.258 \\
    26 & \hspace{2mm} SWIFT J0349.2$-$1159 \hspace{2mm} & 1ES 0347$-$121 & 57.3467 & $-$11.9908 & y* & 5.29 & 3.6 \hspace{000mm} & 0.18 & 45.51 & 20.55 & \ref{Tartarus} & & BL Lac & & & 1.210 \\
    27 & \hspace{2mm} SWIFT J0350.1$-$5019 \hspace{2mm} & PGC 13946 & 57.5990 & $-$50.3099 & y* & 5.99 & 2.9 \hspace{000mm} & 0.036492 & 43.95 & 22.72 & \ref{SwiftXRT} & & Galaxy & \ref{not_obs} & 11.68 & \\
    28 & \hspace{2mm} SWIFT J0356.9$-$4041 \hspace{2mm} & 2MASX J03565655$-$4041453 & 59.2356 & $-$40.6960 & y* & 5.22 & 2.4 \hspace{000mm} & 0.0747 & 44.51 & 22.52 & \ref{SwiftXRT} & & Sy 1.9 & \ref{6dF} & 13.27 & 0.007 \\
    29 & \hspace{2mm} SWIFT J0407.4+0339 \hspace{2mm} & 3C 105 & 61.8186 & 3.7071 & y~ & 4.01 & 3.4 \hspace{000mm} & 0.089 & 44.83 & 23.43 & \ref{SwiftXRT} & & Sy 2 & & 15.16 & \\
    30 & \hspace{2mm} SWIFT J0418.3+3800 \hspace{2mm} & 3C 111.0 & 64.5887 & 38.0266 & & 13.41 & 12.5 \hspace{000mm} & 0.0485 & 44.84 & 21.98 & \ref{Tartarus} & & Sy 1 & & 13.63 & 0.398 \\
    31 & \hspace{2mm} SWIFT J0426.2$-$5711 \hspace{2mm} & 1H 0419$-$577 & 66.5035 & $-$57.2001 & y* & 5.49 & 2.9 \hspace{000mm} & 0.104 & 44.91 & 19.52 & \ref{Tartarus} & & Sy 1 & & & 4.563 \\
    32 & \hspace{2mm} SWIFT J0433.0+0521$^l$  \hspace{2mm} & 3C 120 & 68.2962 & 5.3543 & y* & 13.15 & 11.2 \hspace{000mm} & 0.03301 & 44.45 & 21.19 & \ref{Tartarus} & & Sy 1 & & 11.69 & 2.174 \\
    33 & \hspace{2mm} SWIFT J0444.1+2813 \hspace{2mm} & 2MASX J04440903+2813003 & 71.0376 & 28.2168 & & 7.15 & 7.6 \hspace{000mm} & 0.01127 & 43.33 & 22.72 & \ref{SwiftXRT} & & Sy 2 & & 10.88 & \\
    34 & \hspace{2mm} SWIFT J0451.4$-$0346$^l$  \hspace{2mm} & MCG $-$01$-$13$-$025 & 72.9230 & $-$3.8094 & y* & 5.62 & 4.5 \hspace{000mm} & 0.015894 & 43.41 & 20.62 & \ref{Gallo06} & & Sy 1.2 & & 11.14 & 0.281 \\
    35 & \hspace{2mm} SWIFT J0452.2+4933 \hspace{2mm} & 1RXS J045205.0+493248 & 73.0208 & 49.5459 & & 7.59 & 5.6 \hspace{000mm} & 0.029 & 44.04 & 21.65 & \ref{SwiftXRT} & & Sy 1 & & 12.26 & 0.590 \\
    36 & \hspace{2mm} SWIFT J0505.8$-$2351 \hspace{2mm} & XSS J05054$-$2348 & 76.4405 & $-$23.8539 & y* & 11.26 & 6.1 \hspace{000mm} & 0.035043 & 44.24 & 22.69 & \ref{SwiftXRT} & & Sy 2 & & 13.77 & 0.009 \\
    37 & \hspace{2mm} SWIFT J0510.7+1629 \hspace{2mm} & 4U 0517+17 & 77.6896 & 16.4988 & & 7.12 & 7.8 \hspace{000mm} & 0.017879 & 43.75 & & & & Sy 1.5 & & & 0.670 \\
    38 & \hspace{2mm} SWIFT J0516.2$-$0009$^l$  \hspace{2mm} & Ark 120 & 79.0476 & $-$0.1498 & y* & 7.12 & 5.3 \hspace{000mm} & 0.032296 & 44.11 & 20.30 & \ref{Lutz04} & & Sy 1 & & 11.26 & 2.120 \\
    39 & \hspace{2mm} SWIFT J0501.9$-$3239 \hspace{2mm} & ESO 362$-$G018 & 79.8992 & $-$32.6578 & y* & 10.49 & 5.1 \hspace{000mm} & 0.012642 & 43.26 & 20.25 & & y & Sy 1.5 & & 11.10 & 0.060 \\
    40 & \hspace{2mm} SWIFT J0519.5$-$4545 \hspace{2mm} & PICTOR A & 79.9570 & $-$45.7790 & y~ & 4.23 & 2.2 \hspace{000mm} & 0.035058 & 43.80 & 21.00 & \ref{Tartarus} & & Sy 1/Liner & & 13.63 & 0.626 \\
    41 & \hspace{2mm} SWIFT J0519.5$-$3140 \hspace{2mm} & PKS 0521$-$365 & 80.7416 & $-$36.4586 & y* & 6.02 & 2.8 \hspace{000mm} & 0.05534 & 44.31 & 21.11 & \ref{Tartarus} & & BL Lac & & 12.50 & 0.883 \\
    42 & \hspace{2mm} SWIFT J0538.8$-$4405 \hspace{2mm} & PKS 0537$-$441 & 84.7098 & $-$44.0858 & y* & 5.79 & 3.1 \hspace{000mm} & 0.8904 & 47.09 & 20.54 & \ref{BeppoSax} & & BL Lac & & 13.45 & 0.178 \\
    43 & \hspace{2mm} SWIFT J0539.9$-$2839 \hspace{2mm} & [HB89] 0537$-$286 & 84.9762 & $-$28.6655 & y~ & 4.27 & 2.5 \hspace{000mm} & 3.104 & 48.32 & 20.77 & \ref{Tartarus} & & Blazar & & & 0.092 \\
    44 & \hspace{2mm} SWIFT J0550.7$-$3212 \hspace{2mm} & PKS 0548$-$322 & 87.6699 & $-$32.2716 & y* & 7.39 & 4.4 \hspace{000mm} & 0.069 & 44.70 & 21.50 & \ref{Tartarus} & & BL Lac & & 13.59 & 2.533 \\
    45 & \hspace{2mm} SWIFT J0552.2$-$0727 \hspace{2mm} & NGC 2110 & 88.0474 & $-$7.4562 & y* & 32.46 & 25.6 \hspace{000mm} & 0.007789 & 43.54 & 22.57 & \ref{Tartarus} & & Sy 2 & & 9.26 & 0.010 \\
    46 & \hspace{2mm} SWIFT J0554.8+4625 \hspace{2mm} & MCG +08$-$11$-$011 & 88.7234 & 46.4393 & & 11.37 & 11.1 \hspace{000mm} & 0.020484 & 44.02 & 20.30 & \ref{Lutz04} & & Sy 1.5 & & 10.49 & 1.689 \\
    47 & \hspace{2mm} SWIFT J0557.9$-$3822$^l$  \hspace{2mm} & EXO 055620$-$3820.2 & 89.5083 & $-$38.3346 & y* & 9.82 & 5.2 \hspace{000mm} & 0.03387 & 44.14 & 22.23 & \ref{Tartarus} & y & Sy 1 & & 11.86 & 0.105 \\
    48 & \hspace{2mm} SWIFT J0602.2+2829 \hspace{2mm} & IRAS 05589+2828 & 90.5446 & 28.4728 & & 5.08 & 5.6 \hspace{000mm} & 0.033 & 44.15 & 21.57 & \ref{SwiftXRT} & & Sy 1 & & & 0.866 \\
    49 & \hspace{2mm} SWIFT J0601.9$-$8636 \hspace{2mm} & ESO 005$-$ G 004 & 91.4235 & $-$86.6319 & y* & 5.64 & 4.2 \hspace{000mm} & 0.006228 & 42.56 & 23.88 & \ref{SwiftXRT} & & Sy 2\dag & \ref{Ueda07} & 9.53 & \\
    50 & \hspace{2mm} SWIFT J0615.8+7101$^l$  \hspace{2mm} & Mrk 3 & 93.9015 & 71.0375 & y* & 14.27 & 10.1 \hspace{000mm} & 0.013509 & 43.61 & 24.00 & \ref{Matt00} & y & Sy 2 & & 10.03 & 0.061 \\
    51 & \hspace{2mm} SWIFT J0623.9$-$6058 \hspace{2mm} & ESO 121$-$IG 028 & 95.9399 & $-$60.9790 & y* & 4.85 & 2.8 \hspace{000mm} & 0.0403 & 44.03 & 23.20 & & & Sy 2 & \ref{Donzelli00} & 11.63 & 0.011 \\
    52 & \hspace{2mm} SWIFT J0640.4$-$2554 \hspace{2mm} & ESO 490$-$IG026 & 100.0487 & $-$25.8954 & & 5.14 & 3.6 \hspace{000mm} & 0.0248 & 43.71 & 21.48 & \ref{SwiftXRT} & & Sy 1.2 & & 11.09 & 0.273 \\
    53 & \hspace{2mm} SWIFT J0640.1$-$4328 \hspace{2mm} & 2MASX J06403799$-$4321211 & 100.1583 & $-$43.3558 & y~ & 4.51 & 2.8 \hspace{000mm} & & & 23.04 & \ref{SwiftXRT} & & Galaxy\dag & \ref{not_obs} & 14.24 & \\
    54 & \hspace{2mm} SWIFT J0641.3+3257 \hspace{2mm} & 2MASX J06411806+3249313 & 100.3252 & 32.8254 & & 5.51 & 5.5 \hspace{000mm} & 0.047 & 44.46 & 22.98 & \ref{XMM_RM_inprep} & & Sy 2 & \ref{obs} & 14.01 & \\
    55 & \hspace{2mm} SWIFT J0651.9+7426 \hspace{2mm} & Mrk 6 & 103.0510 & 74.4271 & y* & 9.55 & 6.6 \hspace{000mm} & 0.01881 & 43.72 & 23.00 & \ref{Immler03} & y & Sy 1.5 & & 11.07 & 0.062 \\
    56 & \hspace{2mm} SWIFT J0742.5+4948 \hspace{2mm} & Mrk 79 & 115.6367 & 49.8097 & y* & 7.09 & 4.7 \hspace{000mm} & 0.022189 & 43.72 & 20.76 & \ref{XMM} & & Sy 1.2 & & 11.19 & 2.196 \\
    57 & \hspace{2mm} SWIFT J0746.3+2548 \hspace{2mm} & SDSS J074625.87+254902.2 & 116.6078 & 25.8173 & y* & 5.92 & 4.7 \hspace{000mm} & 2.9793 & 48.55 & 22.00 & \ref{Sambruna06} & & Blazar & \ref{Sambruna06} & & 0.032 \\
    58 & \hspace{2mm} SWIFT J0759.8$-$3844 \hspace{2mm} & IGR J07597$-$3842 & 119.9208 & $-$38.7600 & & 7.79 & 5.3 \hspace{000mm} & 0.04 & 44.29 & 21.70 & \ref{SwiftXRT} & & Sy 1.2 & & & \\
    59 & \hspace{2mm} SWIFT J0841.4+7052$^l$  \hspace{2mm} & [HB89] 0836+710 & 130.3515 & 70.8951 & y* & 11.38 & 7.0 \hspace{000mm} & 2.172 & 48.39 & 20.98 & \ref{Tartarus} & & Blazar & & & 0.755 \\
    60 & \hspace{2mm} SWIFT J0902.0+6007 \hspace{2mm} & Mrk 18 & 135.4933 & 60.1517 & y* & 5.35 & 3.1 \hspace{000mm} & 0.011088 & 42.93 & 23.39 & \ref{XMM_RM_inprep} & y & Galaxy & \ref{Sargent70} & 11.50 & \\
    61 & \hspace{2mm} SWIFT J0904.3+5538 \hspace{2mm} & 2MASX J09043699+5536025 & 136.1539 & 55.6007 & y* & 5.21 & 3.4 \hspace{000mm} & 0.037 & 44.03 & 21.89 & \ref{SwiftXRT} & & Sy 1 & & 13.55 & \\
    62 & \hspace{2mm} SWIFT J0911.2+4533 \hspace{2mm} & 2MASX J09112999+4528060 & 137.8749 & 45.4683 & y* & 5.35 & 3.0 \hspace{000mm} & 0.026782 & 43.69 & 23.42 & \ref{SwiftXRT} & & Sy 2 & & 13.18 & \\
    63 & \hspace{2mm} SWIFT J0917.2$-$6221 \hspace{2mm} & IRAS 09149$-$6206 & 139.0371 & $-$62.3249 & & 4.51 & 3.2 \hspace{000mm} & 0.0573 & 44.40 & 22.19 & \ref{SwiftXRT} & & Sy 1 & & & 0.120 \\
    64 & \hspace{2mm} SWIFT J0918.5+0425 \hspace{2mm} & 2MASX J09180027+0425066 & 139.5011 & 4.4184 & y~ & 4.72 & 3.1 \hspace{000mm} & 0.156 & 45.31 & 23.00 & \ref{SwiftXRT} & & QSO 2$^{**}$ & \ref{SDSS} & 14.91 & \\
    65 & \hspace{2mm} SWIFT J0920.8$-$0805 \hspace{2mm} & MCG $-$01$-$24$-$012 & 140.1927 & $-$8.0561 & y* & 6.44 & 4.6 \hspace{000mm} & 0.019644 & 43.60 & 22.80 & \ref{BeppoSax} & & Sy 2 & & 13.18 & \\
    66 & \hspace{2mm} SWIFT J0923.7+2255$^l$  \hspace{2mm} & MCG +04$-$22$-$042 & 140.9292 & 22.9090 & y* & 6.38 & 4.1 \hspace{000mm} & 0.032349 & 43.99 & 20.60 & \ref{SwiftXRT} & & Sy 1.2 & & 11.83 & 1.626 \\
    67 & \hspace{2mm} SWIFT J0925.0+5218$^l$  \hspace{2mm} & Mrk 110 & 141.3036 & 52.2863 & y* & 9.26 & 5.4 \hspace{000mm} & 0.03529 & 44.19 & 20.58 & \ref{Tartarus} & & Sy 1 & & 13.20 & 1.691 \\
    68 & \hspace{2mm} SWIFT J0945.6$-$1420$^l$  \hspace{2mm} & NGC 2992 & 146.4252 & $-$14.3264 & y* & 9.07 & 6.6 \hspace{000mm} & 0.007709 & 42.94 & 22.00 & \ref{Gilli00} & & Sy 2 & & 9.67 & 0.280 \\
    69 & \hspace{2mm} SWIFT J0947.6$-$3057 \hspace{2mm} & MCG $-$05$-$23$-$016 & 146.9173 & $-$30.9489 & y* & 28.67 & 21.9 \hspace{000mm} & 0.008486 & 43.55 & 22.47 & \ref{RXTE} & & Sy 2 & & 10.53 & 0.256 \\
    70 & \hspace{2mm} SWIFT J0959.5$-$2248$^l$  \hspace{2mm} & NGC 3081 & 149.8731 & $-$22.8263 & y* & 11.34 & 8.8 \hspace{000mm} & 0.007956 & 43.09 & 23.52 & \ref{Maiolino98} & & Sy 2 & & 9.91 & 0.008 \\
    71 & \hspace{2mm} SWIFT J1023.5+1952$^l$  \hspace{2mm} & NGC 3227 & 155.8775 & 19.8650 & y* & 22.01 & 12.9 \hspace{000mm} & 0.003859 & 42.63 & 22.80 & \ref{XMM_Gondoin03} & y & Sy 1.5 & & 8.59 & 0.100 \\
    72 & \hspace{2mm} SWIFT J1031.7$-$3451$^l$  \hspace{2mm} & NGC 3281 & 157.9670 & $-$34.8537 & y* & 10.24 & 7.3 \hspace{000mm} & 0.010674 & 43.27 & 24.30 & \ref{BeppoSax} & y & Sy 2 & & 9.31 & 0.012 \\
    73 & \hspace{2mm} SWIFT J1038.8$-$4942 \hspace{2mm} & 2MASX J10384520$-$4946531 & 159.6854 & $-$49.7826 & & 4.86 & 3.3 \hspace{000mm} & 0.06 & 44.46 & 22.17 & \ref{SwiftXRT} & & Sy 1\dag & \ref{Morelli06} & 13.24 & 0.100 \\
    74 & \hspace{2mm} SWIFT J1040.7$-$4619 \hspace{2mm} & LEDA 093974 & 160.0939 & $-$46.4238 & & 4.26 & 3.4 \hspace{000mm} & 0.023923 & 43.64 & 22.96 & \ref{SwiftXRT} & & Sy 2 & & 11.44 & 0.007 \\
    75 & \hspace{2mm} SWIFT J1049.4+2258 \hspace{2mm} & Mrk 417 & 162.3789 & 22.9644 & y* & 6.39 & 3.6 \hspace{000mm} & 0.032756 & 43.95 & 23.60 & \ref{XMM_RM_inprep} & y & Sy 2 & & 12.74 & \\
    76 & \hspace{2mm} SWIFT J1104.4+3812$^l$  \hspace{2mm} & Mrk 421 & 166.1138 & 38.2088 & y* & 14.02 & 6.8 \hspace{000mm} & 0.030021 & 44.15 & 20.30 & \ref{XMM_Perlman05} & & BL Lac & & 11.09 & 16.220 \\
    77 & \hspace{2mm} SWIFT J1106.5+7234$^l$  \hspace{2mm} & NGC 3516 & 166.6979 & 72.5686 & y* & 18.26 & 10.6 \hspace{000mm} & 0.008836 & 43.26 & 21.21 & \ref{Tartarus} & y & Sy 1.5 & & 9.74 & 4.280 \\
    78 & \hspace{2mm} SWIFT J1127.5+1906 \hspace{2mm} & RX J1127.2+1909 & 171.8178 & 19.1556 & y~ & 4.14 & 2.2 \hspace{000mm} & 0.1055 & 44.79 & 21.30 & \ref{SwiftXRT} & & Sy 1.8 & \ref{Veron_Veron01} & & \\
    79 & \hspace{2mm} SWIFT J1139.0$-$3743$^l$  \hspace{2mm} & NGC 3783 & 174.7572 & $-$37.7386 & y* & 20.46 & 16.1 \hspace{000mm} & 0.00973 & 43.53 & 22.47 & \ref{Lutz04} & y & Sy 1 & & 9.83 & 1.130 \\
    80 & \hspace{2mm} SWIFT J1139.1+5913 \hspace{2mm} & SBS 1136+594 & 174.7873 & 59.1985 & y~ & 4.64 & 2.5 \hspace{000mm} & 0.0601 & 44.33 & 19.58 & \ref{SwiftXRT} & & Sy 1.5 & & 14.83 & 0.372 \\
    81 & \hspace{2mm} SWIFT J1143.7+7942 \hspace{2mm} & UGC 06728 & 176.3168 & 79.6815 & y* & 5.88 & 3.7 \hspace{000mm} & 0.006518 & 42.54 & 20.65 & \ref{XMM_RM_inprep} & & Sy 1.2 & & 11.62 & 0.375 \\
    82 & \hspace{2mm} SWIFT J1145.6$-$1819 \hspace{2mm} & 2MASX J11454045$-$1827149 & 176.4186 & $-$18.4543 & y* & 5.26 & 3.9 \hspace{000mm} & 0.032949 & 43.98 & 20.54 & \ref{SwiftXRT} & & Sy 1 & & 13.93 & 3.293 \\
    83 & \hspace{2mm} SWIFT J1200.8+0650 \hspace{2mm} & CGCG 041$-$020 & 180.2413 & 6.8064 & y~ & 4.53 & 2.5 \hspace{000mm} & 0.036045 & 43.88 & 22.83 & \ref{SwiftXRT} & & Sy 2 & \ref{SDSS} & 12.15 & \\
    84 & \hspace{2mm} SWIFT J1200.2$-$5350 \hspace{2mm} & IGR J12026$-$5349 & 180.6985 & $-$53.8355 & & 5.37 & 4.0 \hspace{000mm} & 0.027966 & 43.86 & 22.34 & & & Sy 2 & & 11.48 & 0.026 \\
    85 & \hspace{2mm} SWIFT J1203.0+4433 \hspace{2mm} & NGC 4051 & 180.7900 & 44.5313 & y* & 9.01 & 4.6 \hspace{000mm} & 0.002335 & 41.74 & 20.47 & \ref{Tartarus} & y & Sy 1.5 & & 8.58 & 3.918 \\
    86 & \hspace{2mm} SWIFT J1204.5+2019 \hspace{2mm} & ARK 347 & 181.1237 & 20.3162 & y~ & 4.39 & 2.3 \hspace{000mm} & 0.02244 & 43.42 & 23.20 & \ref{SwiftXRT} & & Sy 2 & & 11.76 & 0.004 \\
    87 & \hspace{2mm} SWIFT J1206.2+5243 \hspace{2mm} & NGC 4102 & 181.5963 & 52.7109 & y* & 5.00 & 2.4 \hspace{000mm} & 0.002823 & 41.62 & 20.94 & \ref{Chandra} & & Liner & & 8.76 & \\
    88 & \hspace{2mm} SWIFT J1209.4+4340$^l$  \hspace{2mm} & NGC 4138 & 182.3741 & 43.6853 & y~ & 4.53 & 2.1 \hspace{000mm} & 0.002962 & 41.62 & 22.90 & \ref{Risalti99} & & Sy 1.9 & & 9.90 & \\
    89 & \hspace{2mm} SWIFT J1210.5+3924$^l$  \hspace{2mm} & NGC 4151 & 182.6358 & 39.4057 & y* & 74.10 & 37.4 \hspace{000mm} & 0.003319 & 42.96 & 22.48 & \ref{Cappi06} & y & Sy 1.5 & & 8.50 & 0.651 \\
    90 & \hspace{2mm} SWIFT J1218.5+2952 \hspace{2mm} & Mrk 766 & 184.6105 & 29.8129 & y~ & 4.60 & 2.3 \hspace{000mm} & 0.012929 & 42.94 & 21.72 & \ref{Tartarus} & & Sy 1.5 & & 11.10 & 4.710 \\
    91 & \hspace{2mm} SWIFT J1225.8+1240$^l$  \hspace{2mm} & NGC 4388 & 186.4448 & 12.6621 & y* & 45.63 & 25.3 \hspace{000mm} & 0.008419 & 43.60 & 23.63 & \ref{Lutz04} & y & Sy 2 & & 8.98 & 0.516 \\
    92 & \hspace{2mm} SWIFT J1202.5+3332 \hspace{2mm} & NGC 4395 & 186.4538 & 33.5468 & y* & 5.05 & 2.6 \hspace{000mm} & 0.001064 & 40.81 & 22.30 & & y & Sy 1.9 & & 10.66 & \\
    93 & \hspace{2mm} SWIFT J1229.1+0202$^l$  \hspace{2mm} & 3C 273 & 187.2779 & 2.0524 & y* & 44.58 & 26.2 \hspace{000mm} & 0.15834 & 46.25 & 20.54 & \ref{Tartarus} & & Blazar & & 11.69 & 7.905 \\
    94 & \hspace{2mm} SWIFT J1235.6$-$3954$^l$  \hspace{2mm} & NGC 4507 & 188.9026 & $-$39.9093 & y* & 23.56 & 19.3 \hspace{000mm} & 0.011802 & 43.78 & 23.46 & \ref{Lutz04} & y & Sy 2 & & 9.93 & 0.032 \\
    95 & \hspace{2mm} SWIFT J1238.9$-$2720 \hspace{2mm} & ESO 506$-$G027 & 189.7275 & $-$27.3078 & y* & 16.87 & 13.2 \hspace{000mm} & 0.025024 & 44.28 & 23.60 & \ref{SwiftXRT} & y & Sy 2 & \ref{Morelli06} & 11.14 & \\
    96 & \hspace{2mm} SWIFT J1239.3$-$1611 \hspace{2mm} & XSS J12389$-$1614 & 189.7763 & $-$16.1799 & y* & 8.57 & 5.8 \hspace{000mm} & 0.036675 & 44.26 & 22.48 & \ref{SwiftXRT} & & Sy 2 & \ref{Masetti06} & 11.48 & \\
    97 & \hspace{2mm} SWIFT J1239.6$-$0519$^l$  \hspace{2mm} & NGC 4593 & 189.9142 & $-$5.3442 & y* & 14.62 & 9.1 \hspace{000mm} & 0.009 & 43.21 & 20.30 & \ref{Lutz04} & y & Sy 1\dag & & 8.96 & 1.429 \\
    98 & \hspace{2mm} SWIFT J1241.6$-$5748 \hspace{2mm} & WKK 1263 & 190.3572 & $-$57.8343 & & 4.09 & 2.8 \hspace{000mm} & 0.02443 & 43.58 & 21.50 & \ref{XMM_RM_inprep} & & Sy 2\dag & & 12.29 & 0.614 \\
    99 & \hspace{2mm} SWIFT J1256.2$-$0551 \hspace{2mm} & 3C 279 & 194.0465 & $-$5.7893 & y* & 5.47 & 3.2 \hspace{000mm} & 0.5362 & 46.57 & 20.41 & \ref{Tartarus} & & Blazar & & 19.90 & 0.400 \\
    100 & \hspace{2mm} SWIFT J1303.8+5345 \hspace{2mm} & SBS 1301+540 & 195.9978 & 53.7917 & y* & 4.82 & 2.5 \hspace{000mm} & 0.02988 & 43.72 & 20.60 & \ref{SwiftXRT} & & Sy 1 & \ref{Burenin06} & 13.43 & 0.059 \\
    101 & \hspace{2mm} SWIFT J1305.4$-$4928 \hspace{2mm} & NGC 4945 & 196.3645 & $-$49.4682 & & 24.48 & 19.4 \hspace{000mm} & 0.001878 & 42.18 & 24.60 & \ref{Lutz04} & & Sy 2\dag & & 5.60 & 0.085 \\
    102 & \hspace{2mm} SWIFT J1309.2+1139 \hspace{2mm} & NGC 4992 & 197.3040 & 11.6459 & y* & 8.45 & 4.7 \hspace{000mm} & 0.025137 & 43.83 & 23.39 & \ref{XMM_RM_inprep} & y & Galaxy & \ref{SDSS_no_agn_lines} & 11.23 & \\
    103 & \hspace{2mm} SWIFT J1322.2$-$1641$^l$  \hspace{2mm} & MCG $-$03$-$34$-$064 & 200.6019 & $-$16.7286 & y* & 6.53 & 4.7 \hspace{000mm} & 0.016541 & 43.46 & 23.59 & \ref{Risalti99} & y & Sy 1.8 & & 10.80 & \\
    104 & \hspace{2mm} SWIFT J1325.4$-$4301$^l$  \hspace{2mm} & Cen A & 201.3650 & $-$43.0192 & y* & 93.44 & 74.8 \hspace{000mm} & 0.001825 & 42.74 & 22.74 & \ref{Tartarus} & y & Sy 2 & & 4.98 & 0.411 \\
    105 & \hspace{2mm} SWIFT J1335.8$-$3416 \hspace{2mm} & MCG $-$06$-$30$-$015 & 203.9741 & $-$34.2956 & y* & 9.26 & 7.5 \hspace{000mm} & 0.007749 & 43.00 & 21.67 & \ref{Tartarus} & y & Sy 1.2 & & 10.87 & 2.496 \\
    106 & \hspace{2mm} SWIFT J1338.2+0433 \hspace{2mm} & NGC 5252 & 204.5665 & 4.5426 & y* & 10.52 & 6.6 \hspace{000mm} & 0.022975 & 43.90 & 25.82 & \ref{Tartarus} & y & Sy 1.9 & & 10.89 & \\
    107 & \hspace{2mm} SWIFT J1347.4$-$6033 \hspace{2mm} & 4U 1344$-$60 & 206.8500 & $-$60.6400 & & 8.93 & 7.0 \hspace{000mm} & 0.012879 & 45.49 & 22.37 & \ref{ASCA_RM_inprep} & & Sy 1.5 & & & \\
    108 & \hspace{2mm} SWIFT J1349.3$-$3018$^l$  \hspace{2mm} & IC 4329A & 207.3304 & $-$30.3096 & y* & 33.62 & 30.0 \hspace{000mm} & 0.016054 & 44.24 & 21.65 & \ref{Tartarus} & & Sy 1.2 & & 10.24 & 2.960 \\
    109 & \hspace{2mm} SWIFT J1352.8+6917$^l$  \hspace{2mm} & Mrk 279 & 208.2644 & 69.3082 & y* & 8.67 & 4.4 \hspace{000mm} & 0.030451 & 43.97 & 20.53 & \ref{Tartarus} & & Sy 1.5 & & 11.43 & 2.809 \\
    110 & \hspace{2mm} SWIFT J1413.2$-$0312$^l$  \hspace{2mm} & NGC 5506 & 213.3119 & $-$3.2075 & y* & 30.36 & 23.6 \hspace{000mm} & 0.006181 & 43.30 & 22.53 & \ref{Lutz04} & & Sy 1.9 & & 9.71 & 0.110 \\
    111 & \hspace{2mm} SWIFT J1417.7+2539 \hspace{2mm} & 1E 1415+259 & 214.4862 & 25.7240 & y* & 4.92 & 3.1 \hspace{000mm} & 0.237 & 45.71 & 20.72 & \ref{SwiftXRT} & & BL Lac & \ref{Giommi05} & & 1.710 \\
    112 & \hspace{2mm} SWIFT J1417.9+2507 \hspace{2mm} & NGC 5548 & 214.4981 & 25.1368 & y* & 9.11 & 5.8 \hspace{000mm} & 0.01717 & 43.59 & 20.41 & \ref{Tartarus} & y & Sy 1.5 & & 10.64 & 4.950 \\
    113 & \hspace{2mm} SWIFT J1419.0$-$2639 \hspace{2mm} & ESO 511$-$G030 & 214.8434 & $-$26.6447 & y* & 5.73 & 4.7 \hspace{000mm} & 0.02239 & 43.73 & 21.21 & \ref{Tartarus} & & Sy 1 & & 10.79 & 1.221 \\
    114 & \hspace{2mm} SWIFT J1428.7+4234 \hspace{2mm} & 1ES 1426+428 & 217.1361 & 42.6724 & y~ & 4.66 & 2.6 \hspace{000mm} & 0.129 & 45.06 & 21.52 & \ref{Tartarus} & & BL Lac & & & 4.200 \\
    115 & \hspace{2mm} SWIFT J1442.5$-$1715$^l$  \hspace{2mm} & NGC 5728 & 220.5997 & $-$17.2532 & y* & 8.96 & 8.9 \hspace{000mm} & 0.0093 & 43.23 & 23.63 & \ref{Chandra_RM_inprep} & & Sy 2 & & 9.18 & \\
    116 & \hspace{2mm} SWIFT J1504.2+1025 \hspace{2mm} & Mrk 841 & 226.0050 & 10.4378 & y* & 5.56 & 5.1 \hspace{000mm} & 0.036422 & 44.20 & 21.32 & \ref{Tartarus} & y & Sy 1 & & 12.56 & 0.081 \\
    117 & \hspace{2mm} SWIFT J1535.9+5751 \hspace{2mm} & Mrk 290 & 233.9682 & 57.9026 & y~ & 4.66 & 3.0 \hspace{000mm} & 0.029577 & 43.79 & 20.40 & \ref{Tartarus} & & Sy 1 & & 13.04 & 0.885 \\
    118 & \hspace{2mm} SWIFT J1628.1+5145$^l$  \hspace{2mm} & Mrk 1498 & 247.0169 & 51.7754 & y* & 6.13 & 4.5 \hspace{000mm} & 0.0547 & 44.50 & 23.26 & \ref{SwiftXRT} & & Sy 1.9 & & 12.77 & \\
    119 & \hspace{2mm} SWIFT J1648.0$-$3037 \hspace{2mm} & 2MASX J16481523$-$3035037 & 252.0635 & $-$30.5845 & & 6.38 & 8.6 \hspace{000mm} & 0.031 & 44.28 & 21.61 & \ref{SwiftXRT} & & Sy 1 & & 12.56 & 0.149 \\
    120 & \hspace{2mm} SWIFT J1652.9+0223 \hspace{2mm} & NGC 6240 & 253.2454 & 2.4008 & y~ & 4.43 & 4.7 \hspace{000mm} & 0.02448 & 43.81 & 24.34 & \ref{Lutz04} & & Sy 2 & & 10.30 & 0.090 \\
    121 & \hspace{2mm} SWIFT J1654.0+3946 \hspace{2mm} & Mrk 501 & 253.4676 & 39.7602 & y* & 7.63 & 4.9 \hspace{000mm} & 0.03366 & 44.11 & 22.40 & \ref{Tartarus} & y & BL Lac & & 10.67 & 4.122 \\
    122 & \hspace{2mm} SWIFT J1717.1$-$6249 \hspace{2mm} & NGC 6300 & 259.2478 & $-$62.8206 & & 8.76 & 9.1 \hspace{000mm} & 0.003699 & 42.44 & 23.34 & \ref{SwiftXRT} & & Sy 2 & & 7.86 & \\
    123 & \hspace{2mm} SWIFT J1737.5$-$2908 \hspace{2mm} & GRS 1734$-$292 & 264.3512 & $-$29.1800 & & 8.63 & 10.9 \hspace{000mm} & 0.0214 & 44.05 & 21.96 & \ref{ASCA} & & Sy 1 & & & \\
    124 & \hspace{2mm} SWIFT J1745.4+2906 \hspace{2mm} & 1RXS J174538.1+290823 & 266.4094 & 29.1395 & y* & 5.62 & 3.9 \hspace{000mm} & 0.111332 & & 20.67 & \ref{SwiftXRT} & & Sy 1 & \ref{obs} & 13.98 & 0.530 \\
    125 & \hspace{2mm} SWIFT J1835.0+3240 \hspace{2mm} & 3C 382 & 278.7590 & 32.6973 & y* & 10.96 & 8.1 \hspace{000mm} & 0.05787 & 44.81 & 21.13 & \ref{Tartarus} & & Sy 1 & & 11.87 & 2.000 \\
    126 & \hspace{2mm} SWIFT J1838.4$-$6524$^l$  \hspace{2mm} & ESO 103$-$035 & 279.5847 & $-$65.4276 & y* & 9.50 & 9.7 \hspace{000mm} & 0.013286 & 43.58 & 23.17 & \ref{Tartarus} & & Sy 2 & & 11.38 & 0.060 \\
    127 & \hspace{2mm} SWIFT J1842.0+7945$^l$  \hspace{2mm} & 3C 390.3 & 280.5375 & 79.7714 & y* & 17.32 & 10.1 \hspace{000mm} & 0.0561 & 44.88 & 21.03 & \ref{Tartarus} & & Sy 1 & & 12.91 & 0.472 \\
    128 & \hspace{2mm} SWIFT J1930.5+3414 \hspace{2mm} & NVSS J193013+341047 & 292.5554 & 34.1797 & & 5.92 & 3.3 \hspace{000mm} & 0.0629 & 44.50 & 23.20 & \ref{Kennea05} & & Sy 1 & \ref{Halpern06} & 14.24 & \\
    129 & \hspace{2mm} SWIFT J1942.6$-$1024 \hspace{2mm} & NGC 6814 & 295.6694 & $-$10.3235 & y* & 5.68 & 6.2 \hspace{000mm} & 0.005214 & 42.57 & 20.76 & \ref{Reynolds97} & & Sy 1.5 & & 8.66 & 0.034 \\
    130 & \hspace{2mm} SWIFT J1952.4+0237 \hspace{2mm} & 3C 403 & 298.0658 & 2.5068 & & 4.29 & 4.1 \hspace{000mm} & 0.059 & 44.53 & 23.60 & \ref{Kraft05} & & Sy 2 & & 12.53 & \\
    131 & \hspace{2mm} SWIFT J1959.4+4044 \hspace{2mm} & Cyg A & 299.8681 & 40.7339 & & 16.74 & 10.9 \hspace{000mm} & 0.05607 & 44.91 & 23.30 & \ref{Chandra} & & Sy 2 & & 10.61 & 0.947 \\
    132 & \hspace{2mm} SWIFT J1959.6+6507 \hspace{2mm} & 1ES 1959+650 & 299.9994 & 65.1485 & y* & 6.68 & 4.1 \hspace{000mm} & 0.047 & 44.33 & 21.11 & \ref{XMM} & & BL Lac & & 12.54 & 2.653 \\
    133 & \hspace{2mm} SWIFT J2009.0$-$6103 \hspace{2mm} & NGC 6860 & 302.1954 & $-$61.1002 & y* & 5.08 & 4.9 \hspace{000mm} & 0.014884 & 43.39 & 21.75 & \ref{SwiftXRT} & y & Sy 1 & & 10.68 & 0.566 \\
    134 & \hspace{2mm} SWIFT J2028.5+2543a\hspace{2mm} & MCG +04$-$48$-$002 & 307.1463 & 25.7336 & & 9.05 & 6.1 \hspace{000mm} & 0.0139   & 43.42 & 23.60 & \ref{SwiftXRT}      & y & Sy 2 & & 11.23 & \\
    135 & \hspace{2mm} SWIFT J2028.5+2543b\hspace{2mm} & NGC  6921          & 307.1203 & 25.7234 & & 9.05 & 6.1 \hspace{000mm} & 0.014467 & 43.45 & 23.96 & \ref{Winter_inprep} & y & Sy 2 & & 10.01 & \\
    136 & \hspace{2mm} SWIFT J2042.3+7507$^l$  \hspace{2mm} & 4C +74.26 & 310.6554 & 75.1340 & y* & 8.52 & 5.0 \hspace{000mm} & 0.104 & 45.14 & 21.25 & \ref{Ballantyne05} & y & Sy 1 & \ref{Brinkmann98} & & 0.588 \\
    137 & \hspace{2mm} SWIFT J2044.2$-$1045$^l$  \hspace{2mm} & Mrk 509 & 311.0406 & $-$10.7235 & y* & 8.36 & 9.7 \hspace{000mm} & 0.0344 & 44.43 & 20.70 & \ref{Tartarus} & y & Sy 1.2 & & 11.58 & 3.850 \\
    138 & \hspace{2mm} SWIFT J2052.0$-$5704$^l$  \hspace{2mm} & IC 5063 & 313.0097 & $-$57.0688 & y* & 7.90 & 7.1 \hspace{000mm} & 0.011348 & 43.31 & 23.28 & \ref{Tartarus} & y & Sy 2 & & 11.10 & 0.010 \\
    139 & \hspace{2mm} SWIFT J2114.4+8206 \hspace{2mm} & 2MASX J21140128+8204483 & 318.5049 & 82.0801 & y* & 5.86 & 3.6 \hspace{000mm} & 0.084 & 44.80 & 21.11 & \ref{SwiftXRT} & & Sy 1\dag & & 13.17 & 0.460 \\
    140 & \hspace{2mm} SWIFT J2124.6+5057 \hspace{2mm} & IGR J21247+5058 & 321.1589 & 50.9828 & & 21.74 & 13.9 \hspace{000mm} & 0.02 & 44.10 & 22.39 & \ref{SwiftXRT} & & Sy 1 & \ref{Molina06} & & 0.026 \\
    141 & \hspace{2mm} SWIFT J2127.4+5654 \hspace{2mm} & IGR J21277+5656 & 321.9413 & 56.9429 & & 4.21 & 2.7 \hspace{000mm} & 0.0147 & 43.12 & 21.98 & \ref{SwiftXRT} & & Sy 1 & \ref{Masetti06} & & 0.310 \\
    142 & \hspace{2mm} SWIFT J2156.1+4728 \hspace{2mm} & RX J2135.9+4728 & 323.9792 & 47.4731 & & 4.48 & 2.9 \hspace{000mm} & 0.025 & 43.61 & 21.78 & \ref{SwiftXRT} & & Sy 1 & & 12.79 & 0.124 \\
    143 & \hspace{2mm} SWIFT J2152.0$-$3030 \hspace{2mm} & PKS 2149$-$306 & 327.9812 & $-$30.4650 & y* & 5.08 & 5.4 \hspace{000mm} & 2.345 & 48.36 & 20.52 & \ref{SwiftXRT} & & Blazar & & & 0.462 \\
    144 & \hspace{2mm} SWIFT J2200.9+1032 \hspace{2mm} & UGC 11871 & 330.1724 & 10.5524 & y~ & 4.52 & 3.9 \hspace{000mm} & 0.026612 & 43.80 & 22.21 & \ref{SwiftXRT} & & Sy 1.9 & & 11.72 & \\
    145 & \hspace{2mm} SWIFT J2201.9$-$3152$^l$  \hspace{2mm} & NGC 7172 & 330.5080 & $-$31.8698 & y* & 12.28 & 12.4 \hspace{000mm} & 0.008683 & 43.32 & 22.89 & \ref{Tartarus} & & Sy 2 & & 9.44 & 0.012 \\
    146 & \hspace{2mm} SWIFT J2209.4$-$4711 \hspace{2mm} & NGC 7213 & 332.3177 & $-$47.1667 & y* & 6.70 & 5.2 \hspace{000mm} & 0.005839 & 42.59 & 20.60 & \ref{Tartarus} & y & Sy 1.5 & & 7.97 & 3.940 \\
    147 & \hspace{2mm} SWIFT J2235.9$-$2602 \hspace{2mm} & NGC 7314 & 338.9426 & $-$26.0502 & y* & 5.24 & 5.7 \hspace{000mm} & 0.00476 & 42.45 & 21.79 & \ref{Tartarus} & y & Sy 1.9\dag & & 9.06 & 0.236 \\
    148 & \hspace{2mm} SWIFT J2235.9+3358 \hspace{2mm} & NGC 7319 & 339.0148 & 33.9757 & y* & 6.23 & 4.1 \hspace{000mm} & 0.022507 & 43.68 & 23.38 & \ref{Chandra_RM_inprep} & y & Sy 2 & & 11.09 & 0.001 \\
    149 & \hspace{2mm} SWIFT J2246.0+3941 \hspace{2mm} & 3C 452 & 341.4532 & 39.6877 & y~ & 4.78 & 3.3 \hspace{000mm} & 0.0811 & 44.73 & 23.43 & \ref{XRT_Evans06} & & Sy 2 & & 13.35 & \\
    150 & \hspace{2mm} SWIFT J2253.9+1608 \hspace{2mm} & 3C 454.3 & 343.4906 & 16.1482 & y* & 21.25 & 19.0 \hspace{000mm} & 0.859 & 47.83 & 20.77 & \ref{Ginga_Lawson97} & & Blazar & & 14.50 & 0.263 \\
    151 & \hspace{2mm} SWIFT J2254.1$-$1734$^l$  \hspace{2mm} & MR 2251$-$178 & 343.5242 & $-$17.5819 & y* & 9.53 & 10.8 \hspace{000mm} & 0.06398 & 45.03 & 20.80 & \ref{Tartarus} & y & Sy 1 & & 12.54 & 1.037 \\
    152 & \hspace{2mm} SWIFT J2303.3+0852 \hspace{2mm} & NGC 7469 & 345.8151 & 8.8740 & y* & 9.35 & 8.3 \hspace{000mm} & 0.016317 & 43.70 & 20.61 & \ref{Tartarus} & & Sy 1.2 & & 10.11 & 1.700 \\
    153 & \hspace{2mm} SWIFT J2304.8$-$0843 \hspace{2mm} & Mrk 926 & 346.1811 & $-$8.6857 & y* & 5.19 & 5.5 \hspace{000mm} & 0.04686 & 44.45 & 21.14 & \ref{Tartarus} & & Sy 1.5 & & 11.84 & 3.530 \\
    154 & \hspace{2mm} SWIFT J2318.4$-$4223$^l$  \hspace{2mm} & NGC 7582 & 349.5979 & $-$42.3706 & y* & 10.24 & 6.7 \hspace{000mm} & 0.005254 & 42.61 & 22.98 & \ref{Tartarus} & y & Sy 2 & & 8.35 & 0.048 \\
\enddata
\tablenotetext{a}{The {\it Swift} name given is based on the source coordinates from the latest analysis of Swift data 
 except that where a name has been previously published it is kept to avoid confusion. }
\tablenotetext{b}{The ID name given is that of the entry in the NED database (except in those few cases there is none). }
\tablenotetext{c}{J2000 coordinates for the identified counterpart. }
\tablenotetext{d}{`y' indicates that the source is at $|b|>15\degr$  and so, if the SNR is also $>4.8\sigma$ (indicated by 'y*'), is 
 included in the quantitative analysis. }
\tablenotetext{e}{BAT fluxes (in units of $10^{-11}$ erg cm$^{-2}$ s$^{-1}$)
           and luminosities are in the band 14$-$195 keV.
                  Distances for luminosity were calculated using the
                 measured redshift and assuming it was due to Hubble flow. 
              Luminosity errors must
 include the error in measured
flux and the error in distance due to the random velocity of galaxies ($\sim500$ km s$^{-1}$). }
\tablenotetext{f}{Reference for the $n_H$ value - see below. }
\tablenotetext{g}{``cmplx$=$y'' indicates that the spectrum differs significantly from a simple power 
  law with absorption and an Fe line.} 
\tablenotetext{h}{This column contains optically derived types. For well studied AGN, 
 the optical type was derived from  V\'eron-Cetty \& V\'eron (2006).
 For the remaining sources, we determined type by examining
 the spectrum from archival data or from
our own observations. The few remaining AGN without an accessible 
 spectrum are flagged (\dag).}
\tablenotetext{i}{Reference for the type and/or $z$, where this is not from NED - see below }
\tablenotetext{j}{{\it ROSAT} flux in counts s$^{-1}$ from the HEASARC database (Schwope et al., 2000). }
\tablenotetext{k}{The $J$ band is better to use than the $K$ band
                  because it is expected to have a better 
sensitivity in detecting local AGN. The colors of hard X-ray
selected AGN  have $J$/$K$ values $\sim1$ at low redshifts,
      and galaxies at low $z$
also have $J$/$K$ $\sim1$ (Watanabe et al 2004).
      The 2MASS survey is more sensitive in $J$
(http://www.ipac.caltech.edu/2mass/releases/second/doc/figures/secvi2af5.gif), 
where it is shown that the survey goes
$\sim1$ mag more sensitive in $J$ than $K$.}
\tablenotetext{l}{Sources detected in the 3 month survey (Markwardt et al. 2005).}
\tablenotetext{**}{We classify 2MASX J09180027+0425066 as a QSO 
 because its luminosity is greater than $10^{44.5}$ ergs cm$^{-2}$ s$^{-1}$,
     and as type II
because of its very strong narrow OIII lines in SSDS.}

 \newcounter{refno} 
 \setcounter{refno}{0} 
\tablerefs{
(\dag\dag ~indicates our interpretation of published spectra) : 
\refstepcounter{refno} \label{SwiftXRT}           [\arabic{refno}]  {\it Swift XRT} 
\refstepcounter{refno} \label{Akylas02}           [\arabic{refno}]     {\it XTE}  / Akylas  et al. (2002) 
 \refstepcounter{refno} \label{Winter_inprep} [\arabic{refno}]     {\it XMM}  / Winter et al.  (2008b) 
\refstepcounter{refno} \label{Lutz04}               [\arabic{refno}]              Lutz et al. ( 2004) 
 \refstepcounter{refno} \label{Sambruna98}  [\arabic{refno}]          Sambruna et al. (1998) 
  \refstepcounter{refno} \label{ASCA_RM_inprep}   [\arabic{refno}]            {\it     ASCA} / Mushotzky et al.  in preparation 
   \refstepcounter{refno} \label{Gallo06} 
[\arabic{refno}]           Gallo et al. (2006) 
  \refstepcounter{refno} \label{Tartarus} 
[\arabic{refno}]            Tartarus   database 
   \refstepcounter{refno} \label{XMM_RM_inprep} 
[\arabic{refno}]       {\it      XMM}  /  Mushotzky et al. in preparation 
 \refstepcounter{refno} \label{EXOSAT_RM_inprep} 
[\arabic{refno}]      {\it  EXOSAT}  /  Mushotzky et al. in preparation 
 \refstepcounter{refno} \label{Bassani99} 
[\arabic{refno}]       Bassani et al. (1999) 
  \refstepcounter{refno} \label{ROSAT_RM_inprep} 
[\arabic{refno}]     {\it  ROSAT}  /  Mushotzky et al. in preparation 
  \refstepcounter{refno} \label{XMM} 
[\arabic{refno}]           {\it XMM} 
  \refstepcounter{refno} \label{BeppoSax} 
[\arabic{refno}]          {\it BeppoSax} 
   \refstepcounter{refno} \label{Matt00} 
[\arabic{refno}]             Matt et al. (2000) 
   \refstepcounter{refno} \label{Immler03} 
[\arabic{refno}]              Immler et al. (2003) 
   \refstepcounter{refno} \label{Chandra_RM_inprep} 
[\arabic{refno}]         {\it Chandra} / Mushotzky et al. in preparation 
   \refstepcounter{refno} \label{noXRTcounterpart} 
[\arabic{refno}]          no XRT obvious counterpart 
   \refstepcounter{refno} \label{Sambruna06} 
[\arabic{refno}]           Sambruna et al. (2006) 
   \refstepcounter{refno} \label{Gilli00} 
[\arabic{refno}]           Gilli et al. (2000) 
   \refstepcounter{refno} \label{RXTE} 
[\arabic{refno}]           {\it RXTE} 
   \refstepcounter{refno} \label{Maiolino98} 
[\arabic{refno}]           Maiolino et al. (1998) 
   \refstepcounter{refno} \label{XMM_Gondoin03} 
[\arabic{refno}]            {\it XMM} / Gondoin et al. (2003) 
   \refstepcounter{refno} \label{BeppoSax_Vignali02} 
[\arabic{refno}]             {\it BeppoSax} / Vignali,  \& Comastri  (2002) 
   \refstepcounter{refno} \label{XMM_Perlman05} 
[\arabic{refno}]      {\it XMM} / Perlman  et al. (2005) 
   \refstepcounter{refno} \label{Chandra} 
[\arabic{refno}]             {\it Chandra} 
   \refstepcounter{refno} \label{Cappi06} 
[\arabic{refno}]     Cappi  et al. (2006) 
   \refstepcounter{refno} \label{Risalti99} 
[\arabic{refno}]    Risaliti  et al. (1999) 
    \refstepcounter{refno} \label{XRT_no_z} 
[\arabic{refno}]         XRT /  {\it Swift} source/highly absorbed spectrum, no $z$ 
    \refstepcounter{refno} \label{ASCA} 
[\arabic{refno}]          {\it ASCA} 
    \refstepcounter{refno} \label{Kennea05} 
[\arabic{refno}]          Kennea et al. (2005) 
     \refstepcounter{refno} \label{Reynolds97} 
[\arabic{refno}]         Reynolds (1997) 
    \refstepcounter{refno} \label{Kraft05} 
[\arabic{refno}]         Kraft et al. (2005) 
    \refstepcounter{refno} \label{Ballantyne05}             [\arabic{refno}]       Ballantyne  (2005) 
     \refstepcounter{refno} \label{XRT_Evans06}          [\arabic{refno}]      XRT / Evans et al.(2006) 
    \refstepcounter{refno} \label{Ginga_Lawson97}     [\arabic{refno}]       {\it Ginga} / Lawson \& Turner  (1997) 
     \refstepcounter{refno} \label{Veron_Veron01}       [\arabic{refno}]       V\'eron-Cetty \& V\'eron (2001) 
     \refstepcounter{refno} \label{Lewis03}                     [\arabic{refno}]       Lewis, Eracleous \& {Sambruna} (2003) 
   \refstepcounter{refno} \label{obs_no_agn_lines}    [\arabic{refno}]    Observed - no AGN lines  (Winter et al. 2008a) 
   \refstepcounter{refno}  \label{not_obs}        [\arabic{refno}]   No spectrum available 
   \refstepcounter{refno} \label{Schoenmakers98}      [\arabic{refno}]       Schoenmakers  et al. (1998) \dag\dag 
   \refstepcounter{refno} \label{6dF}                    [\arabic{refno}]       6dF \dag\dag 
   \refstepcounter{refno} \label{Ueda07}             [\arabic{refno}]      Ueda et al. (2007) 
     \refstepcounter{refno} \label{Donzelli00}     [\arabic{refno}]       Donzelli (2000) \dag\dag 
     \refstepcounter{refno} \label{obs}                  [\arabic{refno}]   Winter et al. (2008a) 
    \refstepcounter{refno} \label{Sargent70}       [\arabic{refno}]   Sargent (1970) 
   \refstepcounter{refno} \label{SDSS}                 [\arabic{refno}]       SDSS  \dag\dag
   \refstepcounter{refno} \label{Morelli06}           [\arabic{refno}]       Morelli et al. (2006) 
   \refstepcounter{refno} \label{Masetti06}                    [\arabic{refno}]      Masetti et al. (2006b) 
   \refstepcounter{refno} \label{Burenin06}                   [\arabic{refno}]      Burenin (2006) 
   \refstepcounter{refno} \label{SDSS_no_agn_lines}   [\arabic{refno}]       SDSS - no AGN lines 
   \refstepcounter{refno} \label{Giommi05}                   [\arabic{refno}]    Giommi et al.  (2005) 
   \refstepcounter{refno} \label{Halpern06}                    [\arabic{refno}]    Halpern   (2006) 
   \refstepcounter{refno} \label{Brinkmann98}              [\arabic{refno}]     Brinkmann et al.  (1998) 
   \refstepcounter{refno} \label{Molina06}                     [\arabic{refno}]       Molina et al. (2006) 
   \refstepcounter{refno} \label{Bikmaev06}                 [\arabic{refno}]       Bikmaev et al. (2006) 
}
\end{deluxetable}

\clearpage
\end{landscape}

\vfil\eject
\setlength{\voffset}{0mm}
\begin{deluxetable}{lccccc}
\tablewidth{0pt}
\tablecaption{Comparison of fits to the AGN luminosity function\label{lum_fun_tab}}
\tablehead{
\colhead{ } & 
\colhead{Energy} &
\colhead{ } &
\colhead{ } &
\multicolumn{2}{c}{$L_*$ (ergs s$^{-1}$)}   \\ 
\colhead{Reference} &
\colhead{band} &
\colhead{$a$} & 
\colhead{$b$} &
\colhead{ } &
\colhead{ } \\
\multicolumn{6}{l}{$\log L_{14-195}$ (erg s$^{-1})=44$} \\
\colhead{ } &
\colhead{(keV)} &
\colhead{ } &
\colhead{ } &
\colhead{Native band} & 
\colhead{14$-$195 keV}
}
\startdata
This work                                         & 14$-$195   & $0.84^{+0.16}_{-0.22}$       &  $2.55^{+0.43}_{-0.30}$                   &                                                 & $43.85\pm0.26$      \\
\noalign{\smallskip}
Beckmann et al. (2006b)               &  20$-$40    & $0.80\pm0.15$                      &  $2.11\pm0.22$                    &  $43.38\pm 0.35$                & $43.99\pm 0.35$     \\
Sazonov et al. (2007)                     &    17$-$60   & $0.76\:^{+0.18}_{-0.20}$    &  $2.28\:^{+0.28}_{-0.22}$   & $43.40\pm{0.28}$                & $43.74\pm 0.28$      \\
Barger et al. (2005)                         &   2$-$8         & $0.42\pm0.06$                     &  $2.2\pm0.5$                       &  $44.11\pm 0.08$                 & $44.54\pm 0.08$      \\
La Franca et al. (2005)                   &   2$-$10       & $0.97\:^{+0.08}_{-0.10}$   &  $2.36^{+0.13}_{-0.11}$     &  $44.25\pm0.18$                  & $44.61\pm 0.18$       \\
Sazonov and Revnivtsev (2004) & 3$-$20        & $0.88\:^{+0.18}_{-0.20}$    &  $2.24\:^{+0.22}_{-0.18}$   & $43.58\:^{+0.32}_{-0.30}$  & $43.83\:^{+0.32}_{-0.30}$
\enddata
\tablecomments{
Luminosities have been converted to 14$-$195 keV values assuming a low energy slope of 1.7 breaking to 2.0 at 10 keV.  Uncertainties do not take into account the uncertainty in the conversion.\\
La Franca et al. quote a range of solutions; a representative one is used here.\\
The normalization of the BAT AGN luminosity function (A) is 
          $1.8^{+2.7}_{-1.1} \times 10^{-5}$  erg s$^{-1}$ Mpc$^{-3}$ 
       at $\log L$(erg  s$^{-1}$)$=44$.}
\end{deluxetable}

\end{document}